\documentclass[reprint,aps,pra,showpacs,floatfix,twocolumn]{revtex4-1}
\usepackage{graphicx}
\usepackage{dsfont}
\usepackage{epstopdf}

\newcommand{\be}{\begin{equation}}
\newcommand{\ee}{\end{equation}}

\begin{document}

\title{Persistence of unvisited sites in quantum walks on a line}

\author{M. \v Stefa\v n\'ak\email[correspondence to:]{martin.stefanak@fjfi.cvut.cz}}
\affiliation{Department of Physics, Faculty of Nuclear Sciences and Physical Engineering, Czech Technical University in Prague, B\v
rehov\'a 7, 115 19 Praha 1 - Star\'e M\v{e}sto, Czech Republic}

\author{I. Jex}
\affiliation{Department of Physics, Faculty of Nuclear Sciences and Physical Engineering, Czech Technical University in Prague, B\v
rehov\'a 7, 115 19 Praha 1 - Star\'e M\v{e}sto, Czech Republic}


\date{\today}

\begin{abstract}
We analyze the asymptotic scaling of persistence of unvisited sites
for quantum walks on a line. In contrast to the classical random
walk there is no connection between the behaviour of persistence and
the scaling of variance. In particular, we find that for a two-state
quantum walks persistence follows an inverse power-law where the
exponent is determined solely by the coin parameter. Moreover, for a
one-parameter family of three-state quantum walks containing the
Grover walk the scaling of persistence is given by two
contributions. The first is the inverse power-law. The second
contribution to the asymptotic behaviour of persistence is an
exponential decay coming from the trapping nature of the studied
family of quantum walks. In contrast to the two-state walks both the
exponent of the inverse power-law and the decay constant of the
exponential decay depend also on the initial coin state and its coherence. Hence, one can achieve various regimes of persistence by altering the initial
condition, ranging from purely exponential decay to purely inverse
power-law behaviour.
\end{abstract}

\maketitle

\section{Introduction}

Quantum walks \cite{adz,meyer,fg} represent a versatile tool in
quantum information processing with applications ranging from search
algorithms \cite{skw,childs:search:2004,childs:search:2014,vasek:search}, graph
isomorphism testing \cite{gamble,berry,rudiger}, finding
structural anomalies in graphs \cite{reitzner,hillery,cottrell} or perfect state transfer \cite{kendon:qw:pst,wojcik:qw:pst,barr:qw:pst,zhan:qw:pst,gedik:qw:pst}. Moreover, quantum walks were shown to be universal tools for quantum computation \cite{childs,Lovett}.

One fundamental characterization of classical random walks on
infinite lattices \cite{montroll:1964} is recurrence or transience.
Random walk is said to be recurrent when the probability to return
to the starting point at some later time (so-called P\'olya number)
is unity, and transient otherwise. In fact, recurrence ensures that
any lattice point is visited with certainty. P\'olya has shown
\cite{polya} that for unbiased random walks this property depends on
the dimension of the lattice. In particular, random walks are
recurrent in dimensions 1 and 2 and transient on cubic and higher
dimensional lattices. This result originates from the diffusive
behaviour of a classical random walk.

Since measurement has a non-trivial effect on the state of the
quantum system, one has to specify a particular measurement scheme
to extend the concept of recurrence to the domain of quantum walks.
One possibility is to consider a scheme \cite{stef:prl} where the
quantum walk is restarted from the beginning after the measurement,
and in each iteration one additional step is performed. In this way
the effect of measurement on the quantum state is minimized. Within
this measurement scheme the P\'olya number of a quantum walk depends
not only on the dimension of the lattice, but also on the coin
operator which drives the walk, and in some cases also on the
initial coin state \cite{stef:pra}. The ballistic nature of quantum
walks implies that most of them are transient already in dimension
2. However, some quantum walks, such as the Grover walk
\cite{inui:2dgrover,inui:grover1,inui:grover2,watabe:grover,falkner},
show the so-called trapping effect (or localization). This feature
can be employed to construct recurrent quantum walks in arbitrary
dimension \cite{stef:pra}.

Another scheme is to continue with the quantum walk evolution after
the measurement \cite{rec:kiss}. The effect of frequent measurement
is that the quantum walks are transient already on a one-dimensional
lattice, as follows from \cite{ambainis}. Recurrence of quantum
state within this measurement scheme has been analyzed for general
discrete time unitary evolution in \cite{Grunbaum:recurrence}. The
authors have found that the expectation value of the first return
time is quantized, i.e. it is either infinite or an integer. More
recently, it was shown \cite{sinkovicz:recurrence} that this
property is preserved even in iterated open quantum dynamics,
provided that the corresponding superoperator is unital in the
relevant part of the Hilbert space. Moreover, the notion of
monitored recurrence was extended to a finite-dimensional subspace
in \cite{Grunbaum:rec:subspace}. In such case the averaged expected
return time is a rational number.

Persistence describes the probability that a given site remains
unvisited until certain number of steps. As such, it can be viewed
as a complementary event to that of recurrence. For classical random
walks on a line and a plane persistence of any site tends to zero
for large number of steps. In particular, on one-dimensional lattice
persistence obeys an inverse power-law with exponent $1/2$, which
follows in a straightforward way from the diffusive behaviour of a
random walk \cite{chandra:1943}.

In the context of quantum walks persistence was first introduced in
\cite{pers:goswani}. The authors have analyzed persistence for
two-state Hadamard walk on a line within the measurement scheme of
\cite{stef:prl}, i.e. when the quantum walk is restarted after the
measurement. It was found that persistence of any site follows an
inverse power-law with exponent determined numerically as
$\lambda\approx 0.318$. In contrast to the classical case, no clear
connection of the exponent to the spreading properties of the
quantum walk was found.

In the present paper we give analytical explanation of the results
found in \cite{pers:goswani}. The study of persistence is extended
to a one-parameter set of two-state quantum walks on a line. We
confirm that persistence obeys an inverse power-law. The exponent is
determined solely by the parameter of the coin operator. Hence,
there is no connection of the exponent to the scaling of variance
like in the classical random walk. Moreover, we analyze persistence
of sites for a set of three-state quantum walks \cite{stef:cont:def,machida,stef:limit} which involves the familiar
Grover walk \cite{inui:grover1,inui:grover2,falkner} as a special
case. We find that persistence exhibits a more complicated
asymptotic behaviour. In addition to the inverse power-law there is
also an exponential decay which arises from the trapping effect. The analytical results are obtained using the suitable basis of the coin space formed by the eigenvectors of the coin operator \cite{stef:limit}. Both the exponent of the inverse power-law and the decay rate of the exponential decay depend on the coin parameter and, in contrast to the two-state walk, on the initial coin state and its coherence. Hence, it is possible to obtain various regimes of persistence, ranging from pure inverse power-law to pure exponential decay, by choosing different initial
condition. Moreover, we find that for some initial coin states
persistence behave differently for lattice sites on the positive and
negative half-lines.

The paper is organized as follows: In Section~\ref{sec:2} we review
the definition of persistence of site $m$ within a particular
measurement scheme of \cite{stef:prl,pers:goswani}. We provide an
estimate of the asymptotic behaviour of persistence based on the
limit density. Section~\ref{sec:3} is dedicated to the analysis of
persistence in two-state quantum walks. In Section~\ref{sec:4} we
perform similar analysis for a set of three-state quantum walks.
More technical details are left for Appendices~\ref{app:a} and
\ref{app:b}. We conclude and present an outlook in
Section~\ref{sec:5}.

\section{Persistence of Unvisited Sites}
\label{sec:2}

In this Section we briefly introduce persistence of a given site and
provide and estimate of its asymptotic behaviour. We follow the
measurement scheme used in \cite{stef:prl,pers:goswani}, where the
quantum walk is restarted from the beginning after each measurement.
By persistence of a site $m$ we understand the probability that the
particular lattice point remains unvisited until $T$ steps. Since the walk starts at the origin of the lattice we only consider persistence of sites $m\neq 0$. We find
that this probability is given by \cite{pers:goswani}
\begin{equation}
\label{pers:def}
{\cal P}_m(T) = \prod_{t=1}^T (1-p(m,t)),
\end{equation}
where $p(m,t)$ denotes the probability to find the quantum particle at position $m$ after $t$ steps of the quantum walk.

Let us now turn to the approximation of persistence for large $T$. For this purpose we re-write (\ref{pers:def}) in the exponential form
\begin{eqnarray}
\nonumber {\cal P}_m(T) & = & \exp\left(\ln\left(\prod\limits_{t=1}^T (1-p(m,t))\right)\right)\\
\nonumber  & = & \exp\left(\sum_{t=1}^T \ln(1-p(m,t))\right).
\end{eqnarray}
We replace the logarithm by the first order Taylor expansion and arrive at
\begin{equation}
\label{pers:approx1}
{\cal P}_m(T) \approx \exp\left(-\sum_{t=1}^T p(m,t)\right).
\end{equation}
Next, we use the limit density $w(v)$ derived from the weak-limit theorem \cite{Grimmett} to estimate the exact probability $p(m,t)$ by
$$
p(m,t) \approx \frac{1}{t}w\left(\frac{m}{t}\right).
$$
Finally, we estimate the sum in (\ref{pers:approx1}) with an integral
\begin{equation}
\label{pers:I}
{\cal I}_m(T) = \int\limits_{1}^T \frac{1}{t}w\left(\frac{m}{t}\right)dt
\end{equation}
and obtain the approximation of persistence
\begin{equation}
\label{pers:approx}
{\cal P}_m(T) \approx \exp\left(-{\cal I}_m(T)\right).
\end{equation}
In the following we analyze persistence of unvisited sites for two-
and three-state quantum walks on a line. Detailed evaluations of the
integral (\ref{pers:I}) are left for the Appendices.


\section{Two-state walk on a line}
\label{sec:3}

Let us start our analysis with the two-state quantum walk on a line with the coin operator
$$
C(\rho) = \left(
            \begin{array}{cc}
              \rho & \sqrt{1-\rho^2} \\
              \sqrt{1-\rho^2} & -\rho \\
            \end{array}
          \right),\quad 0 < \rho < 1.
$$
The coin parameter $\rho$ determines the speed of propagation of the
wave packet on the line \cite{kempf}. For $\rho=1/\sqrt{2}$ we
obtain the familiar Hadamard walk \cite{ambainis}.

Suppose that the initial coin state of the particle was
$$
|\psi_C\rangle = a|L\rangle + b|R\rangle.
$$
The limiting probability density for the two-state quantum walk is given by \cite{Konno:2002a,Konno:2005}
$$
w(v) = \frac{\frac{\sqrt{1-\rho^2}}{\rho}\left(1- v \Lambda(a,b)\right)}{\pi  \left(1-v^2\right)\sqrt{1-\frac{v^2}{\rho^2}}},
$$
where $\Lambda$ is determined by the initial coin state and the coin parameter
$$
\Lambda(a,b) = |a|^2 - |b|^2 + \frac{\sqrt{1-\rho^2}}{\rho}\left(a\overline{b} + b\overline{a}\right).
$$
Before we turn to the persistence we first simplify the dependence
on the initial coin state $\Lambda$ by changing the basis of the
coin space. Following the idea of \cite{stef:dir:col} we consider
the basis formed by the eigenvectors of the coin operator
\begin{eqnarray}
\label{basis:2state}
\nonumber |\chi^+\rangle & = & \sqrt{\frac{1+\rho}{2}}|L\rangle + \sqrt{\frac{1-\rho}{2}}|R\rangle,\\
|\chi^-\rangle & = & -\sqrt{\frac{1-\rho}{2}}|L\rangle + \sqrt{\frac{1+\rho}{2}}|R\rangle,
\end{eqnarray}
which satisfy the relations
$$
C(\rho) |\chi^\pm\rangle = \pm|\chi^\pm\rangle.
$$
We decompose the initial coin state of the walk into the eigenvector basis as
$$
|\psi_C\rangle = h_+|\chi^+\rangle + h_-|\chi^-\rangle.
$$
From (\ref{basis:2state}) we find that the coefficients of the initial coin state in the standard basis $a$ and $b$ are related to the eigenbasis coefficients $h_\pm$ by
\begin{eqnarray}
\nonumber a & = & \frac{\sqrt{1+\rho}}{\sqrt{2}}h_+ - \frac{\sqrt{1-\rho}}{\sqrt{2}}h^-, \\
\nonumber  b & = & \frac{\sqrt{1-\rho}}{\sqrt{2}}h_+ + \frac{\sqrt{1+\rho}}{\sqrt{2}}h^-.
\end{eqnarray}
In the new basis the factor $\Lambda(a,b)$ becomes
$$
\Lambda(h_+,h_-) = \frac{2|h_+|^2-1}{\rho}.
$$
The asymptotic probability density thus simplifies into
\begin{equation}
\label{had:asymp:dist}
w(v) = \frac{\frac{\sqrt{1-\rho^2}}{\rho}\left(1- \frac{v}{\rho}(2|h_+|^2-1) \right)}{\pi \left(1-v^2\right)\sqrt{1-\frac{v^2}{\rho^2}}}
\end{equation}

Let us now turn to persistence. We leave the details of evaluation of the integral (\ref{pers:approx}) for Appendix~\ref{app:a}. We find that for large $T$ the function ${\cal I}_m(T)$ grows like a logarithm
$$
{\cal I}_m(T) \sim \lambda\ln\left(\frac{T}{|m|}\right),
$$
where the pre-factor reads
\begin{equation}
\label{lambda:2state}
\lambda = \frac{\sqrt{1-\rho^2}}{\rho\pi}.
\end{equation}
Hence, we find that in the asymptotic regime persistence of site $m$ follows an inverse power-law
\begin{equation}
\label{plaw:2state}
{\cal P}_m(T) \sim \left(\frac{T}{|m|}\right)^{-\lambda},
\end{equation}
The exponent $\lambda$ is independent of the initial coin state. It
is determined solely by the coin operator, i.e. by the value of
$\rho$. Note that for $\rho=1/\sqrt{2}$, i.e. the Hadamard walk, we
find that $\lambda = \frac{1}{\pi}\approx 0.318$, which is in
agreement with the numerical result obtained in \cite{pers:goswani}.

\begin{figure}[h!]
\includegraphics[width=0.45\textwidth]{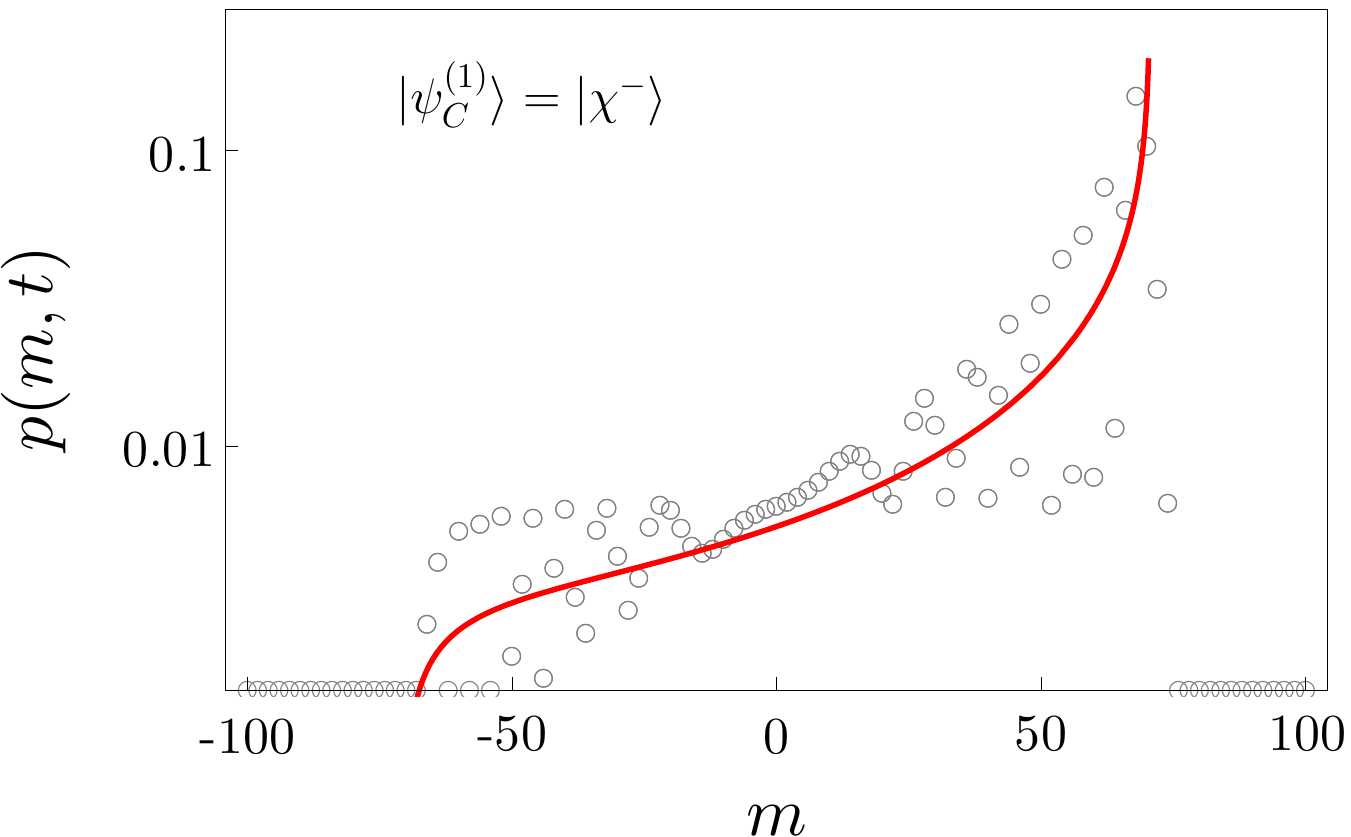}\vspace{12pt}
\includegraphics[width=0.45\textwidth]{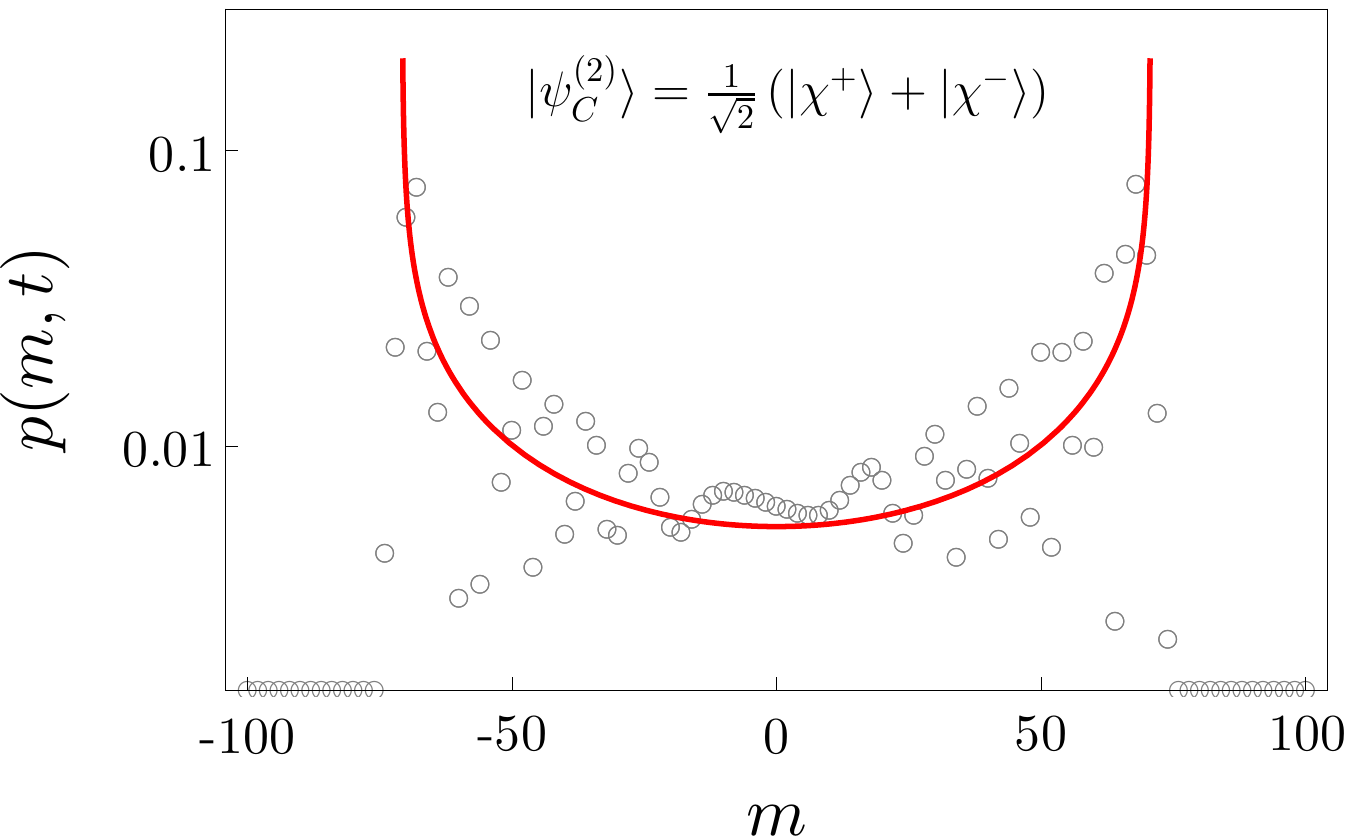}\vspace{12pt}
\includegraphics[width=0.45\textwidth]{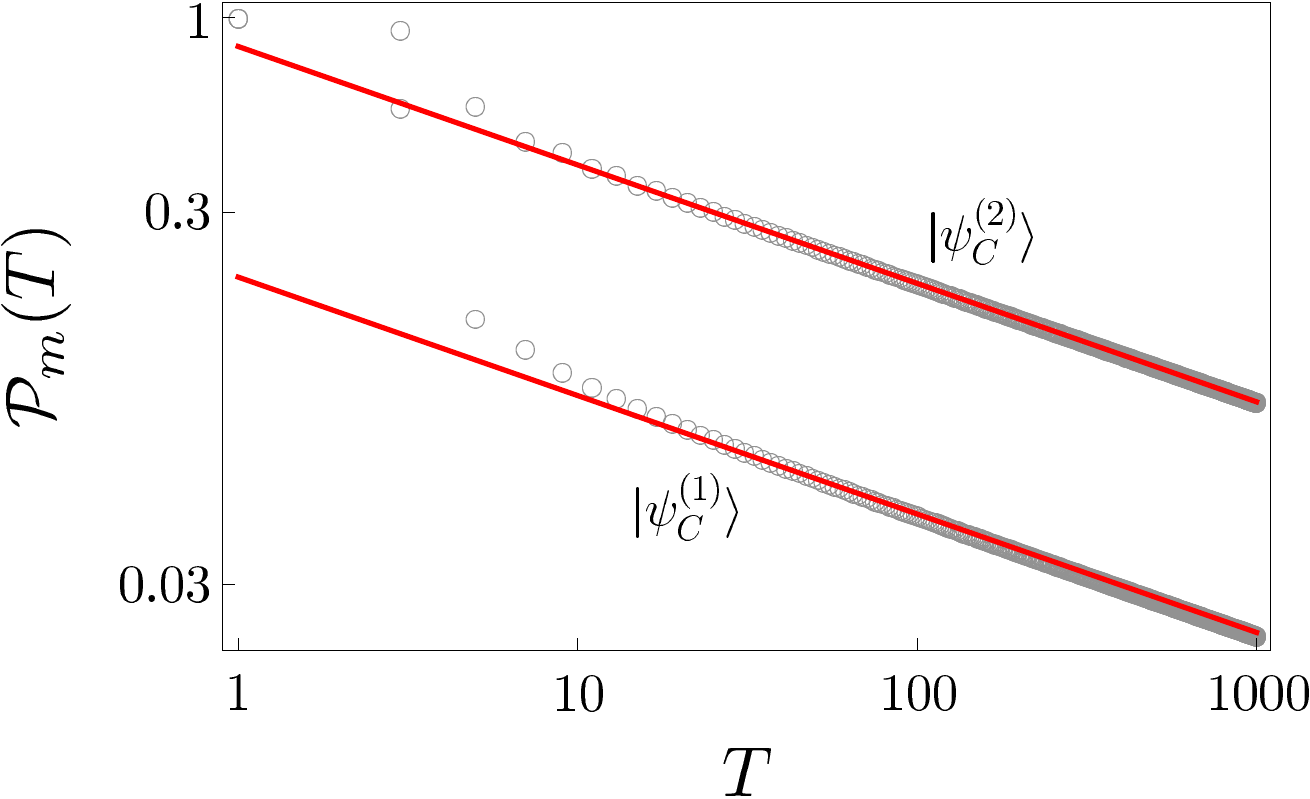}
\caption{Probability density and persistence in dependence on the choice of the initial coin state for the Hadamard walk ($\rho=1/\sqrt{2}$). The first two plots show the probability distribution of the Hadamard walk after $t=100$ steps for two different initial coin states $|\psi_C^{(1,2)}\rangle$ on a semi-log scale. In the lower plot we display persistence of site $m=2$ as a function of the number of steps $T$ on a log-log scale. Despite the differences in the probability distributions the asymptotic scaling of persistence is independent of the initial state, in accordance with (\ref{plaw:2state}).}
\label{fig1}
\end{figure}

\begin{figure}[h!]
\includegraphics[width=0.45\textwidth]{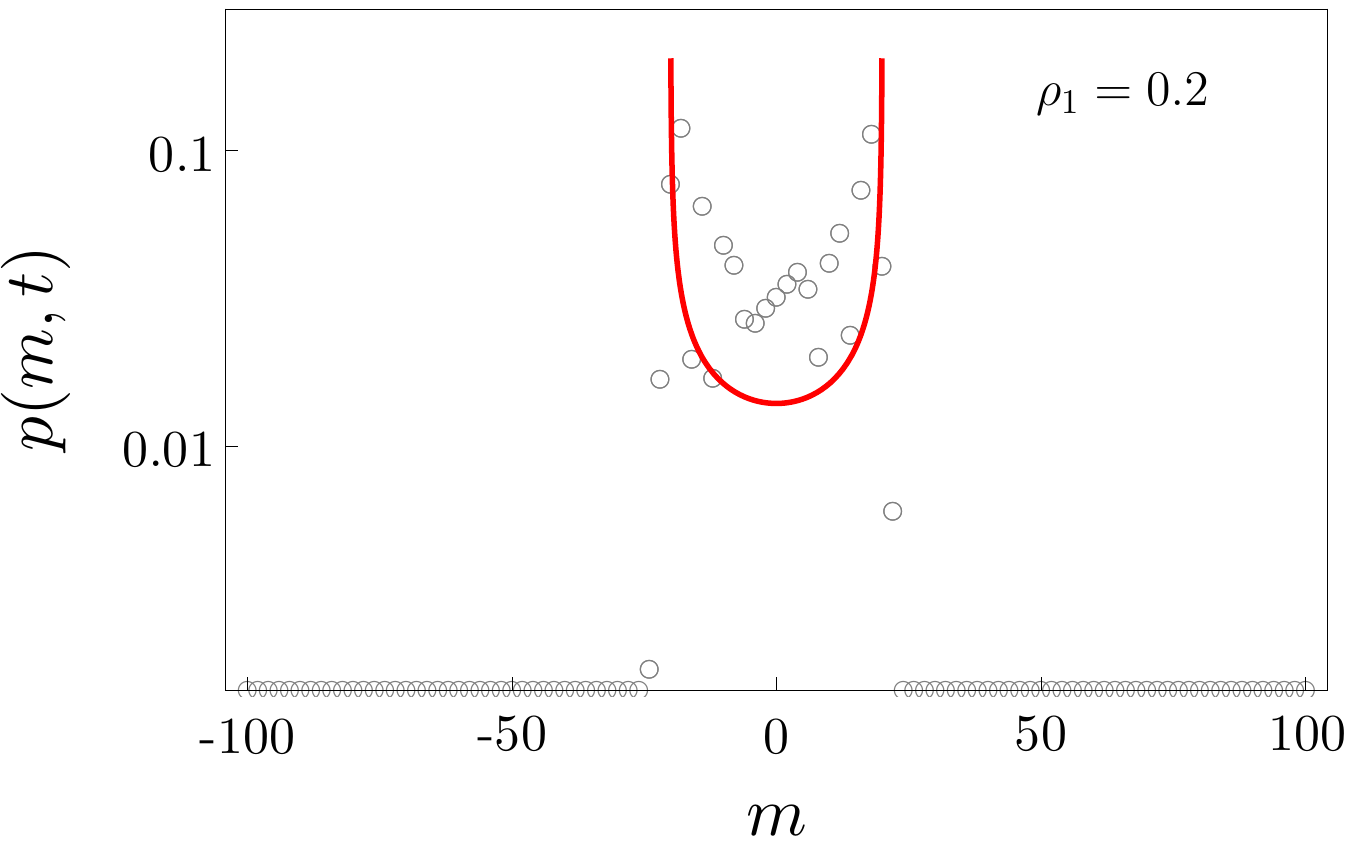}\vspace{12pt}
\includegraphics[width=0.45\textwidth]{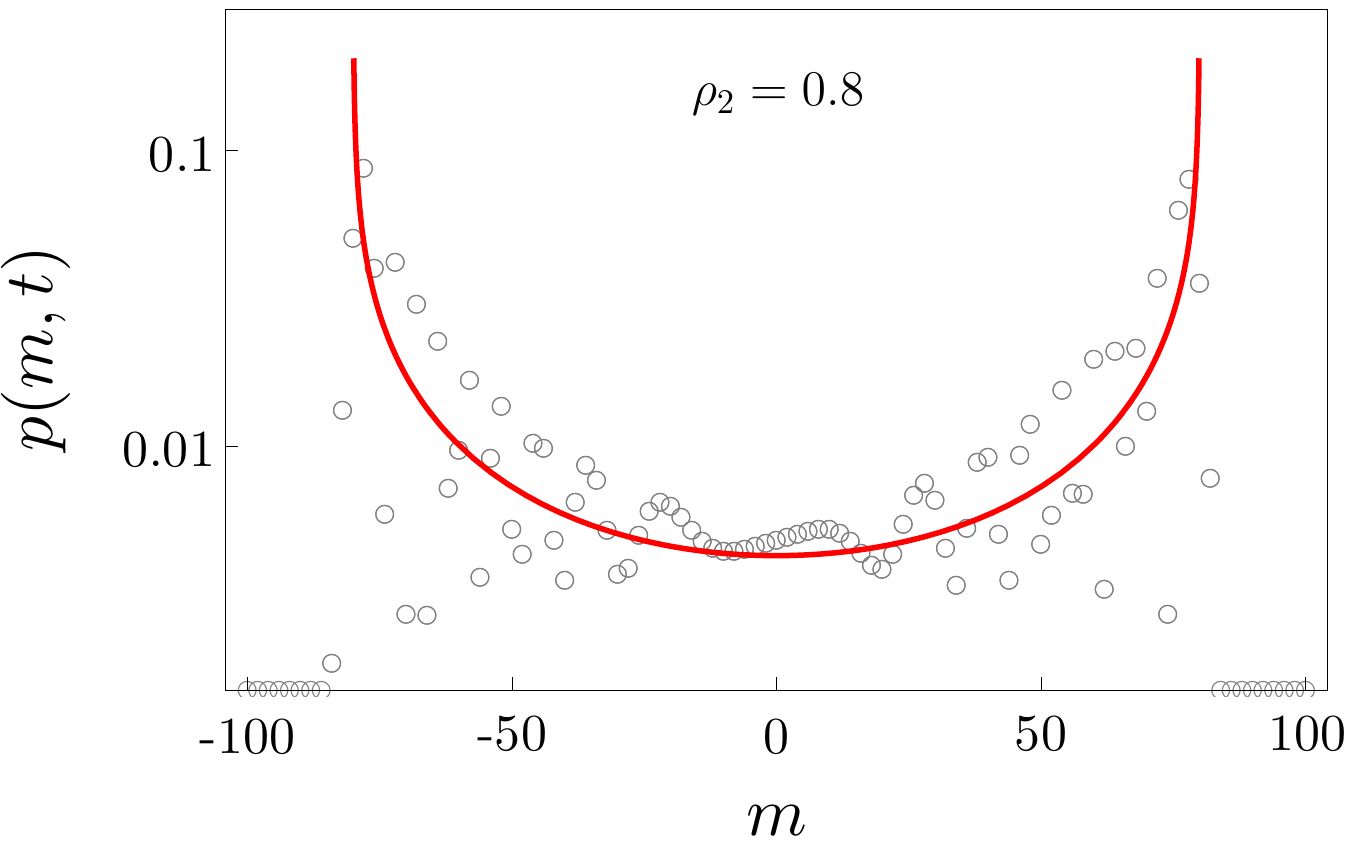}\vspace{12pt}
\includegraphics[width=0.45\textwidth]{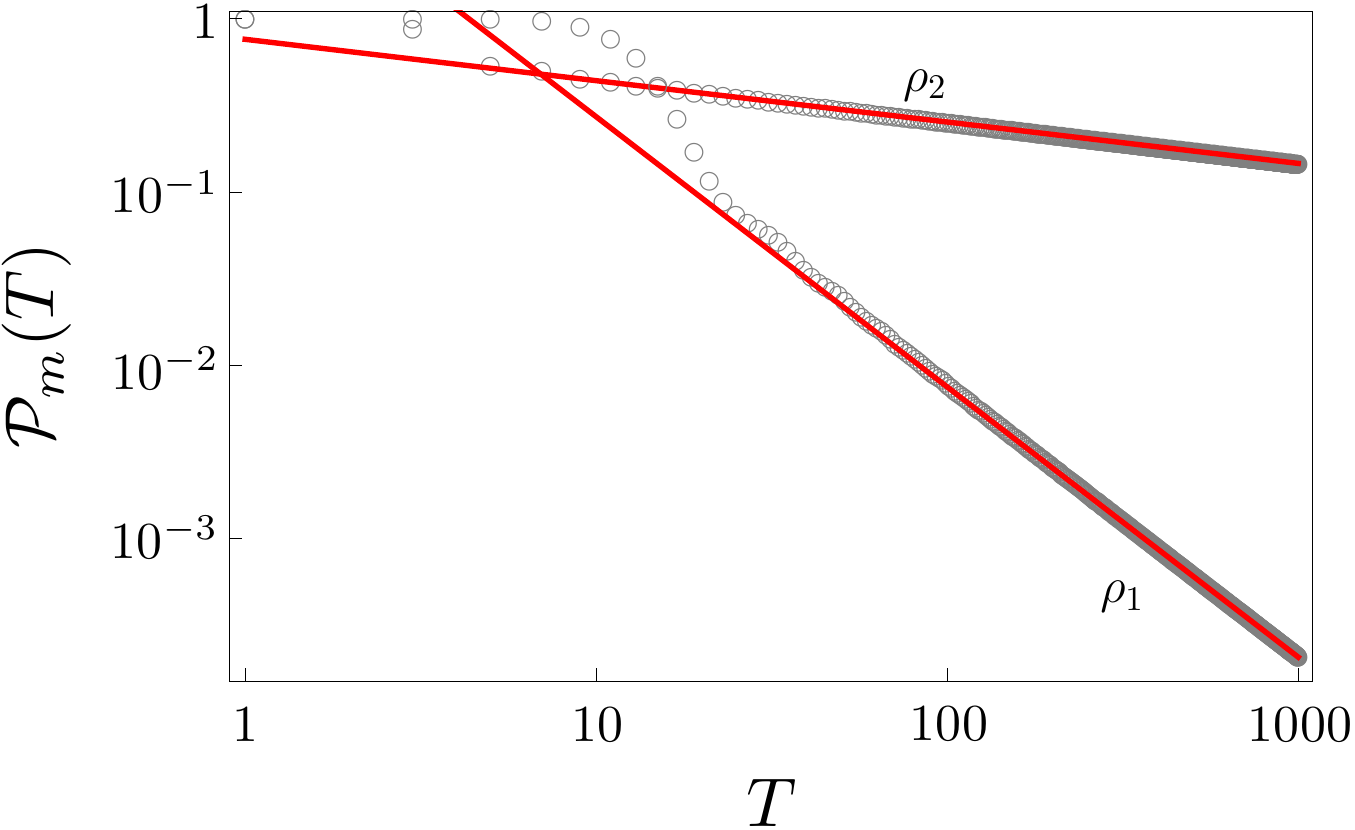}
\caption{Probability density and persistence in dependence on the choice of the coin parameter $\rho$. The first two plots show the probability distribution of the two-state quantum walk on a semi-log scale. In both situations the initial coin state was chosen as $|\psi_C\rangle = \frac{1}{\sqrt{2}}(|\chi^+\rangle + |\chi^-\rangle)$, which leads to symmetric probability distribution. In the upper plot the coin parameter is $\rho_1=0.2$ while in the middle plot we have chosen $\rho_2 = 0.8$. The lower plot shows scaling of persistence of site $m=2$ for different values of $\rho_{1,2}$ on a log-log scale. The exponent of the inverse power-law (\ref{plaw:2state}) decreases with increasing value of $\rho$, as predicted by (\ref{lambda:2state}).
}
\label{fig2}
\end{figure}

\begin{figure}[h!]
\includegraphics[width=0.45\textwidth]{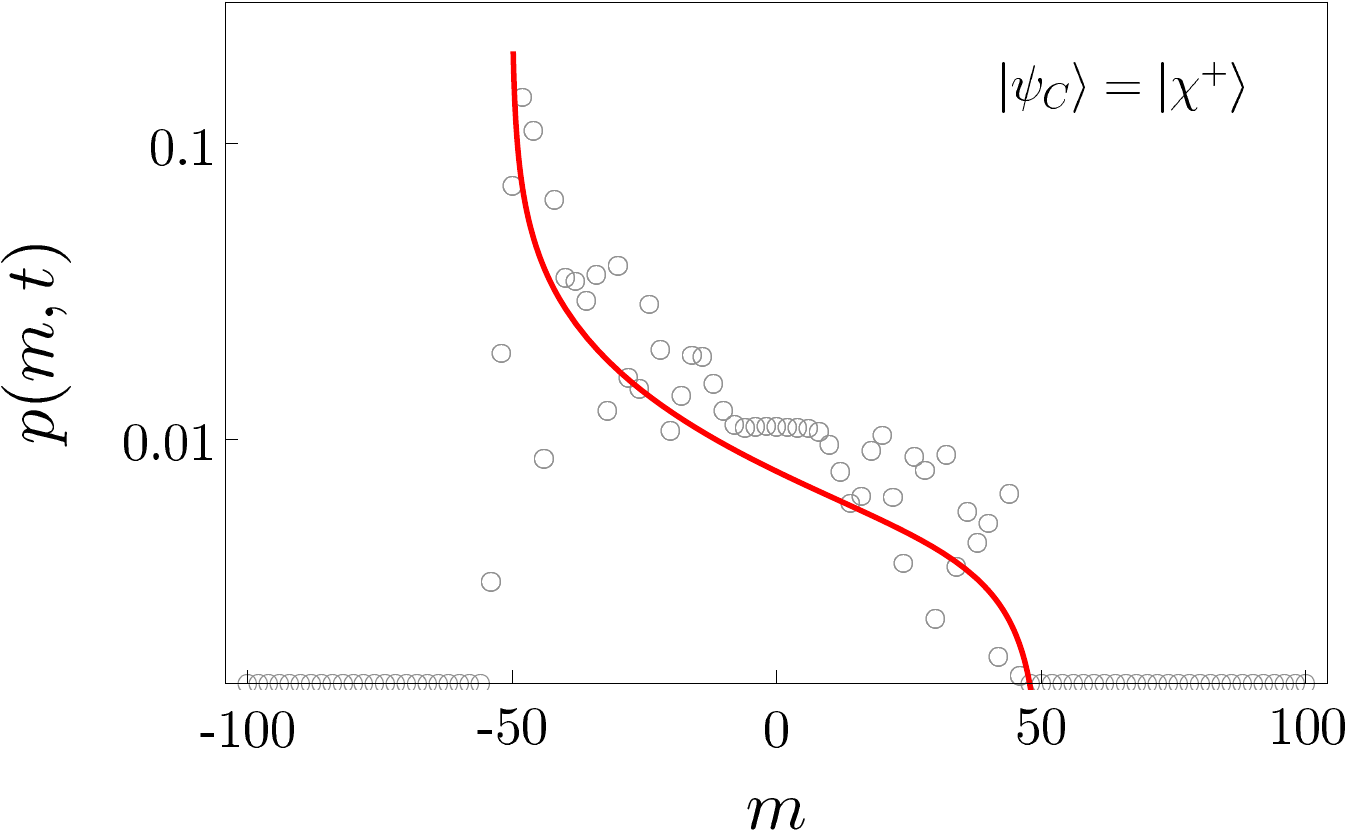}\vspace{12pt}
\includegraphics[width=0.45\textwidth]{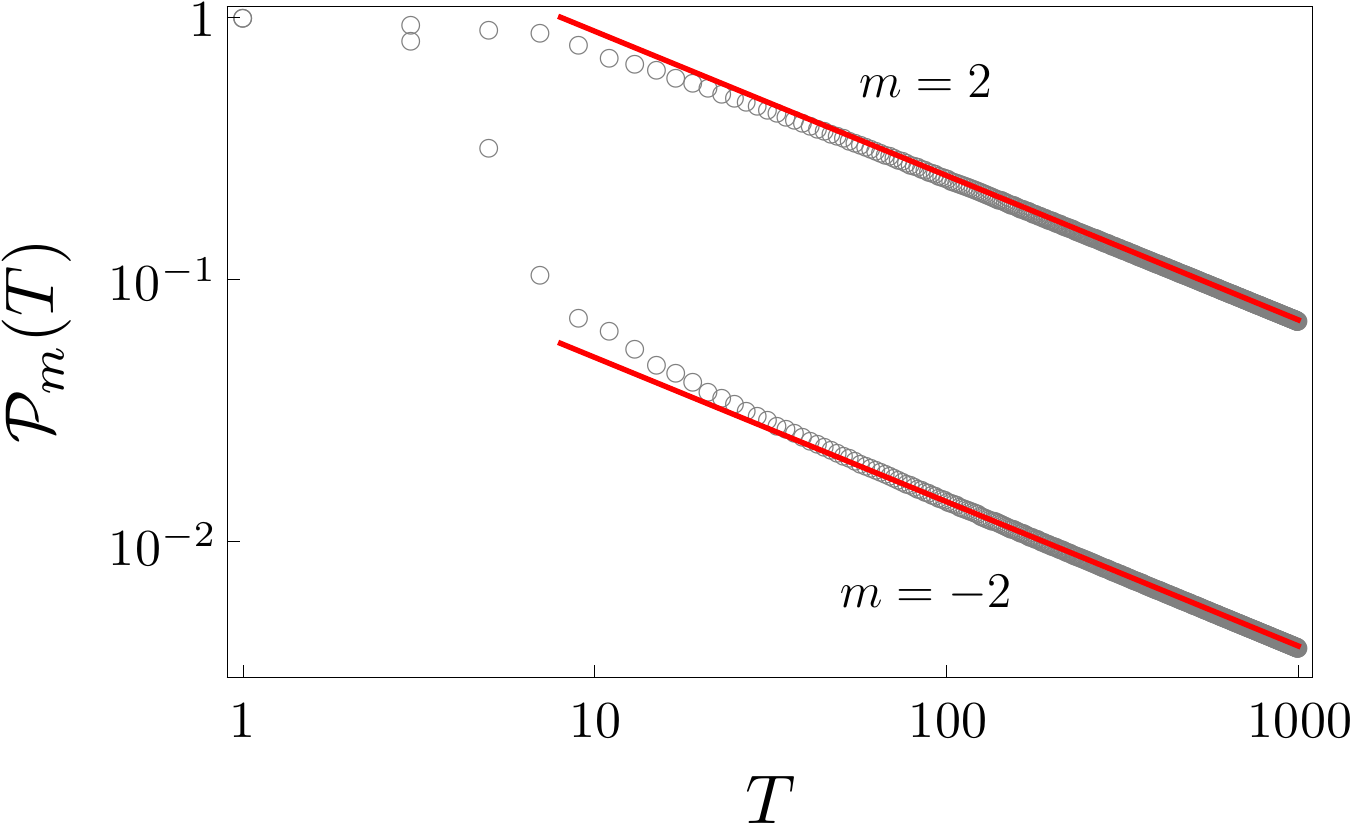}
\caption{The first plot shows the probability distribution for the two-state walk with $\rho=0.5$. The initial coin state is $|\psi_C\rangle = |\chi^+\rangle$, which results in density with only one peak on the left. The lower plot illustrates the behaviour of persistence of sites $m=2$ and $m=-2$. Both curves have the same slope (\ref{lambda:2state}) which is determined solely by the coin parameter $\rho$.}
\label{fig2a}
\end{figure}

Our results are illustrated in Figures~\ref{fig1}-\ref{fig2a}. In
Figure~\ref{fig1} we show the influence of the initial coin state.
In the first two plots we display the probability distribution of
the two-state quantum walk with the coin parameter
$\rho=1/\sqrt{2}$, i.e. the Hadamard walk. In all Figures grey
circles represent the data-points obtained from numerical
simulation. The red curves correspond to the asymptotic probability
density given by (\ref{had:asymp:dist}). For the upper plot we have
chosen the initial coin state $|\psi_C^{(1)}\rangle =
|\chi^-\rangle$. The resulting probability density shows only one
peak on the right. In the middle plot the initial coin state was
chosen according to $|\psi_C^{(2)}\rangle =
\frac{1}{\sqrt{2}}\left(|\chi^+\rangle + |\chi^-\rangle\right)$.
This state leads to a symmetric distribution. Despite the
differences in the probability distributions the persistence shows
the same asymptotic scaling, as we illustrate in the last figure.
Here we display the persistence of site $m=2$ as a function of the
number of steps $T$. To unravel the inverse power-law behavior we
use log-log scale. The grey circles correspond to the numerical
simulation and the red curves show the inverse power-law
(\ref{plaw:2state}).


In Figure~\ref{fig2} we illustrate the influence of the coin
parameter $\rho$. In the first two plots we show the probability
distribution of the two-state quantum walk with the initial coin
state $|\psi_C\rangle = \frac{1}{\sqrt{2}}(|\chi^+\rangle +
|\chi^-\rangle)$. For the upper plot the coin parameter is
$\rho_1=0.2$. In the middle plot we have chosen the coin parameter
$\rho_2 = 0.8$. We see that the coin parameter directly affects the
speed at which the walk spreads through the lattice \cite{kempf}.
The lower plot shows the difference in the scaling of persistence of
site $m=2$ for different values of $\rho$. We use log-log scale to
unravel the scaling of persistence. We find that the exponent of the
inverse power-law decreases with increasing value of $\rho$, in
accordance with (\ref{lambda:2state}).


Finally, Figure~\ref{fig2a} illustrates that the asymptotic
behaviour of persistence is independent of the actual position $m$.
The upper plot displays the probability distribution of the
two-state walk with coin parameter $\rho=0.5$ and the initial coin
state $|\psi_C\rangle = |\chi^+\rangle$. This initial condition
leads to a density which is the most biased towards left, as
indicated by the presence of only one peak. In the lower plot we
show persistence of sites $m=2$ and $m=-2$ on a log-log scale.
Despite the differences in the intermediate regime, the slope of
both curves is the same, in agreement with (\ref{plaw:2state}).


To conclude this Section, we have found that for the two-state
quantum walk on a line persistence of unvisited sites obeys an
inverse power-law (\ref{plaw:2state}) with exponent
(\ref{lambda:2state}) determined only by the coin parameter.


\section{Three-state walk on a line}
\label{sec:4}

Let us now turn to the three-state walk on a line. Here the particle is allowed to move to the left, stay at its position or move to the right. We denote the corresponding orthogonal coin states by $|L\rangle$, $|S\rangle$ and $|R\rangle$. As for the coin operator we consider the one which was studied in \cite{stef:cont:def,machida,stef:limit}. In the standard basis $\left\{|L\rangle,|S\rangle,|R\rangle\right\}$ the coin operator is given by the matrix
\begin{equation}
\label{coin:rho}
C(\rho) = \left(
  \begin{array}{ccc}
    -\rho^2 & \rho\sqrt{2-2\rho^2} & 1-\rho^2 \\
    \rho\sqrt{2-2\rho^2} & 2\rho^2-1 & \rho\sqrt{2-2\rho^2} \\
    1-\rho^2 & \rho\sqrt{2-2\rho^2} & -\rho^2 \\
  \end{array}
\right),
\end{equation}
with parameter $\rho\in (0,1)$. Quantum walks with such a coin
operator represent a one-parameter extension of the familiar
three-state Grover walk \cite{inui:grover1,inui:grover2,falkner},
which corresponds to the choice of $\rho=1/\sqrt{3}$. Indeed, the
results of \cite{stef:limit} have shown that the considered quantum
walks share the same features and the coin factor $\rho$ is a
scaling parameter which determines the rate of spreading of the
three-state quantum walk across the line.

To evaluate persistence of unvisited sites we estimate the exact
probability distribution $p(m,t)$ for large number of steps $t$. In
contrast to the two-state walk, the properties of the probability
distribution are not fully captured by the limit density $w(v)$.
Indeed, the three-state quantum walk leads to the trapping effect
\cite{inui:grover1,inui:grover2,falkner,machida,stef:limit}, which
means that the probability of finding the particle at position $m$
has a non-vanishing limit for $t$ approaching infinity. We denote
the limiting value
$$
\lim\limits_{t\rightarrow\infty} p(m,t) = p_\infty(m),
$$
as the trapping probability. Hence, for large $t$ we approximate the probability to find the particle at position $m$ at time $t$ with the sum
$$
p(m,t) \approx \frac{1}{t}w\left(\frac{m}{t}\right) + p_\infty(m).
$$
The limit density $w(v)$ and the trapping probability $p_\infty(m)$
were analyzed in \cite{machida,stef:limit}. We follow the results of
\cite{stef:limit} since they have simpler form due to the use of a
more suitable basis of the coin space. In particular, the basis of
the coin space was constructed from the eigenvectors of the coin
operator (\ref{coin:rho}) which reads
\begin{eqnarray}
\nonumber |\sigma^+\rangle & = & \sqrt{\frac{1 - \rho^2}{2}}|L\rangle +\rho|S\rangle + \sqrt{\frac{1 - \rho^2}{2}}|R\rangle, \\
\nonumber |\sigma_1^-\rangle & = & \frac{\rho}{\sqrt{2}}|L\rangle -\sqrt{1-\rho^2}|S\rangle + \frac{\rho}{\sqrt{2}}|R\rangle, \\
\nonumber  |\sigma_2^-\rangle & = & \frac{1}{\sqrt{2}}(|L\rangle - |R\rangle).
\end{eqnarray}
The vectors satisfy the eigenvalue equations
$$
C(\rho)|\sigma^+\rangle = |\sigma^+\rangle,\qquad C(\rho)|\sigma_i^-\rangle = -|\sigma_i^-\rangle, i=1,2.
$$
We decompose the initial coin state into the eigenstate basis according to
$$
|\psi_C\rangle = g_+|\sigma^+\rangle + g_1|\sigma_1^-\rangle + g_2|\sigma_2^-\rangle.
$$
The limiting probability density then reads \cite{stef:limit}
\begin{eqnarray}
\label{dist:rho}
w(v) & = & \frac{\sqrt{1-\rho^2}}{\pi(1-v^2)\sqrt{\rho^2-v^2}} \left(1-|g_2|^2\frac{}{} - \right.\\
\nonumber & & \left. - (g_1\overline{g_2} + \overline{g_1}g_2)\frac{v}{\rho} + (|g_2|^2 - |g_+|^2)\frac{v^2}{\rho^2}\right).
\end{eqnarray}
The trapping probability is given by \cite{stef:limit}
\begin{equation}
\label{loc:rho}
p_\infty(m) = \left\{
                \begin{array}{c}
                  \frac{2-2\rho^2}{\rho^4}Q^{2m} |g_+ + g_2|^2 ,\quad m>0 , \\
                   \\
                   \frac{Q}{\rho^2}\left\{|g_+|^2 + (1-\rho^2)|g_2|^2 \right\}, \quad m = 0,\\
                   \\
                  \frac{2-2\rho^2}{\rho^4}Q^{2|m|} |g_+ - g_2|^2 ,\quad m<0 \\
                \end{array}
              \right.
\end{equation}
where $Q$ depends on the coin parameter $\rho$
$$
Q = \frac{2-\rho^2-2\sqrt{1-\rho^2}}{\rho^2}.
$$

Let us estimate the persistence of site $m$. We approximate the sum in (\ref{pers:approx1}) with
$$
\sum_{t=1}^T p(m,t) \approx {\cal I}_m(T) +  \sum_{t = \left\lceil\frac{|m|}{\rho}\right\rceil}^T p_\infty(m),
$$
where ${\cal I}_m(T)$ is defined in (\ref{pers:I}). The sum on the right hand side is trivial
$$
\sum_{t = \left\lceil\frac{|m|}{\rho}\right\rceil}^T p_\infty(m) = \left(T-\left\lceil\frac{|m|}{\rho}\right\rceil\right) p_\infty(m).
$$
Here $\lceil x\rceil$ denotes the ceiling of $x$, i.e. the smallest
integer not less than $x$. The integral ${\cal I}_m(T)$ is evaluated
in Appendix~\ref{app:b}. We find that ${\cal I}_m(T)$ asymptotically
grows like a logarithm
$$
{\cal I}_m(T) \sim \lambda\ln\left(\frac{T}{|m|}\right),
$$
where the pre-factor reads
$$
\label{lambda:3state}
\lambda = \frac{\sqrt{1-\rho^2}}{\pi  \rho} \left(1-\left| g_2\right| ^2\right).
$$
We conclude that for the three-state quantum walk on a line persistence of site $m$ behaves asymptotically like
\begin{equation}
\label{pers:3state}
{\cal P}_m(T) \sim \left(\frac{T}{|m|}\right)^{-\lambda} e^{-p_\infty(m)T}.
\end{equation}
We see that there are two contributions to persistence. Similarly to
the two-state walk there is an inverse power-law.  In addition, the
trapping effect contributes with the exponential decay. However, the
behavior of persistence depends on the initial state, in contrast to
the two-state walk. Indeed, both $\lambda$ and the trapping
probability $p_\infty(m)$ are determined by the initial condition. The exponent $\lambda$ depends only on
the probability $\vert g_2\vert^2$ to find the initial coin state $|\psi_C\rangle$ in the eigenstate $|\sigma_2^-\rangle$. On the other hand, the rate of the exponential decay is determined by the interference of the amplitudes $g_+$ and $g_2$. In the following we discuss various initial conditions to illustrate
our result.

\begin{figure}[h!]
\includegraphics[width=0.45\textwidth]{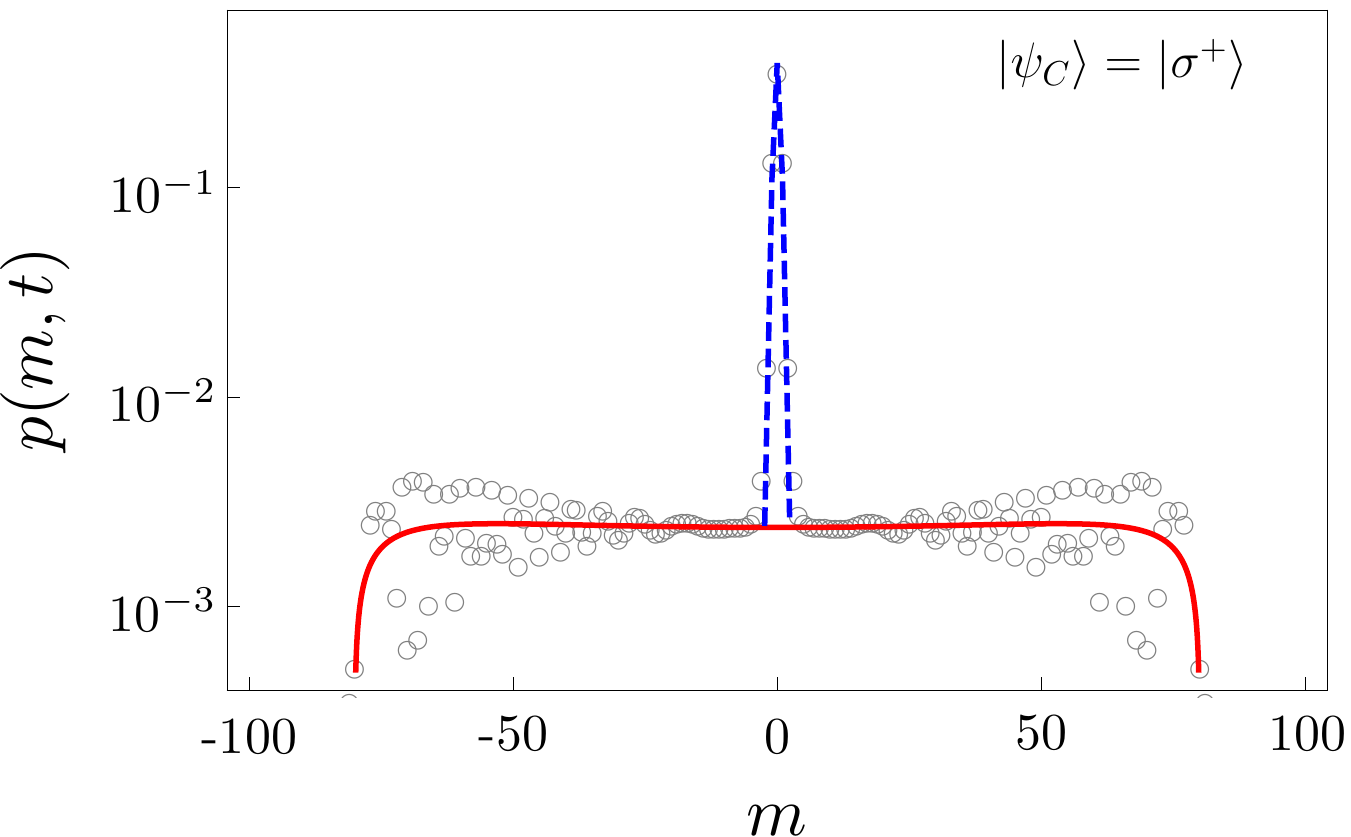}\vspace{12pt}
\includegraphics[width=0.45\textwidth]{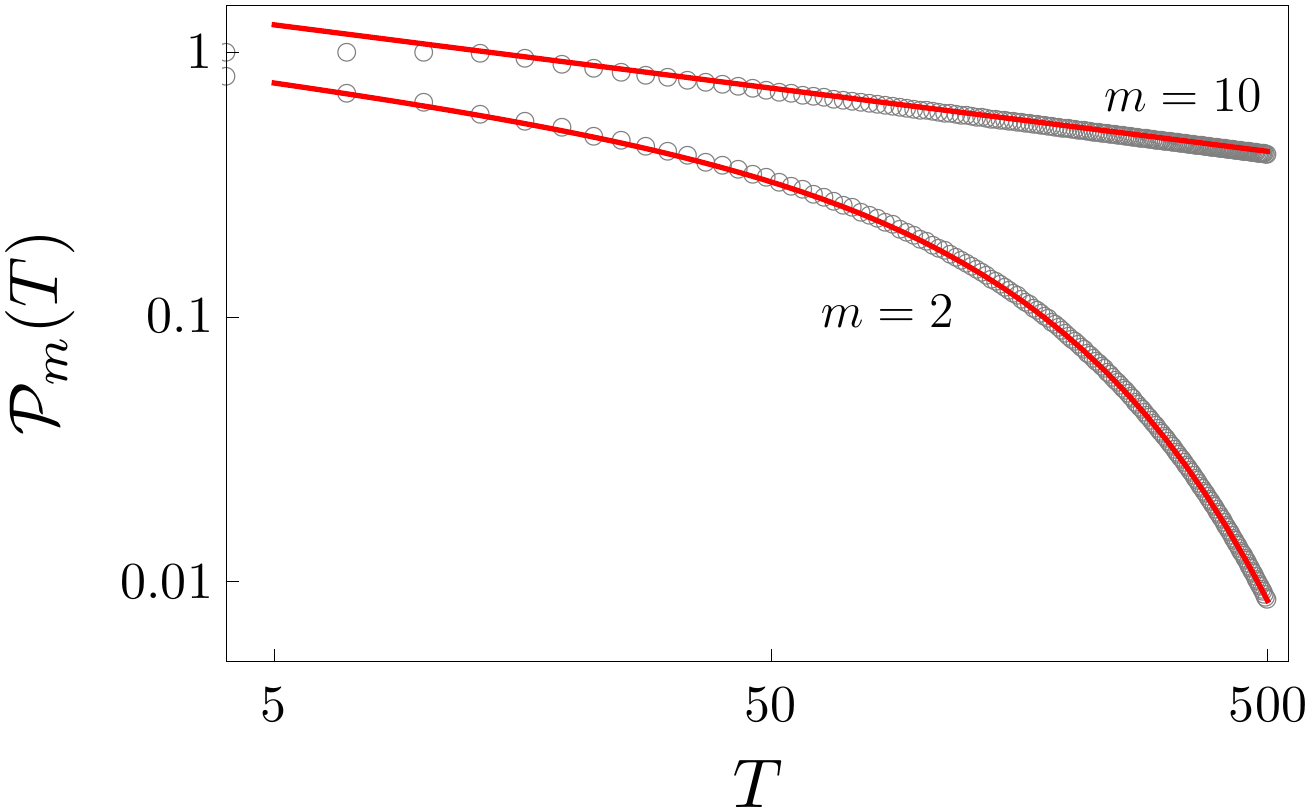}\vspace{12pt}
\includegraphics[width=0.45\textwidth]{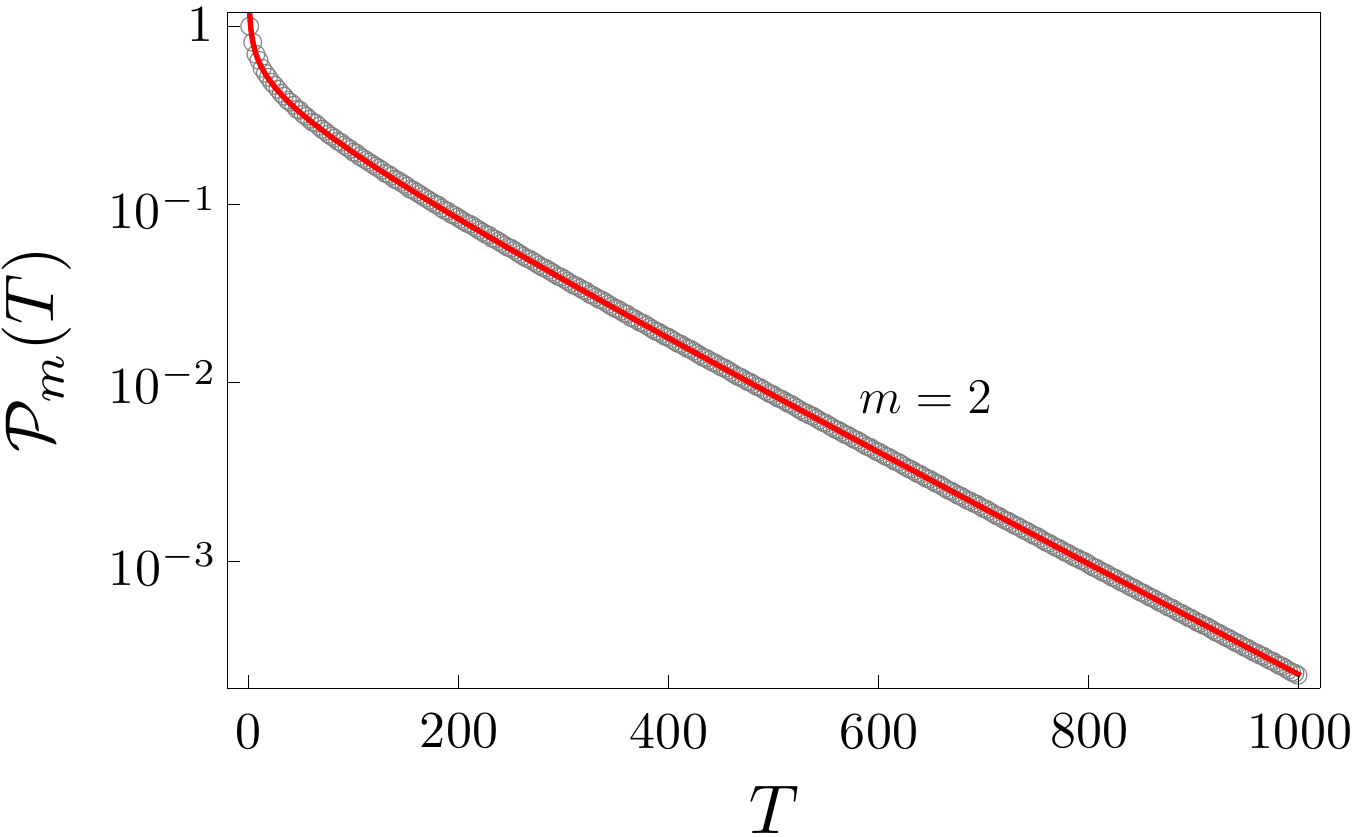}
\caption{Probability distribution and persistence for the three-state walk with $\rho=0.8$ starting with the coin state $|\psi_C\rangle = |\sigma^+\rangle$.
In the first plot we show the probability distribution after $t=100$ steps. The second plot displays persistence (\ref{pers:gp}) of sites $m=2$ and $m=10$ on the log-log scale. For $m=2$ the decay of persistence is faster than the inverse power-law. The deviation is due to the exponential decay which starts to play a dominant role for large $T$. We do not observe this effect for $m=10$, since the decay constant decreases exponentially with the distance from the origin. The third plot, which shows persistence of site $m=2$ on the log-scale, confirms that ${\cal P}_m(T)$ decays exponentially.}
\label{fig3}
\end{figure}

Let us first consider the initial coin state $|\psi_C\rangle =
|\sigma^+\rangle$. In such a case the general formula
(\ref{pers:3state}) for the asymptotic behaviour of persistence
turns into
\begin{equation}
\label{pers:gp}
{\cal P}_m^{(g_+)}(T) \sim \left(\frac{T}{|m|}\right)^{-\lambda} e^{-\gamma(m)T},
\end{equation}
with the exponent given by
\begin{equation}
\label{power:decay}
\lambda = \frac{\sqrt{1-\rho^2}}{\pi\rho},
\end{equation}
and the decay constant
\begin{equation}
\label{decay:const:gp}
\gamma(m) = \frac{2(1-\rho^2)}{\rho^4}Q^{2|m|}.
\end{equation}
We see that both contributions, namely the inverse power-law and the
exponential decay, are present. For illustration of this result, we
show in Figure~\ref{fig3} the probability distribution and
persistence for the three-state walk with the coin parameter
$\rho=0.8$. The first plot displays the probability distribution
after $t=100$ steps. The grey circles corresponds to the numerical
simulation, the red curve depicts the asymptotic probability density
(\ref{dist:rho}) and the blue dashed curve corresponds to the
trapping probability (\ref{loc:rho}). The second plot illustrates
persistence of sites $m=2$ and $m=10$ on the log-log scale. For
$m=2$ the decay of persistence is faster than inverse power-law.
Indeed, for large number of steps the exponential decay starts to
play a dominant role. On the other hand, for $m=10$ we do not
observe any deviation from the inverse power-law at the considered
time-scale. This is due to the fact that the decay constant
(\ref{decay:const:gp}) itself decreases exponentially with the
distance from the origin. The last plot, where we display
persistence of site $m=2$ on the logarithmic scale, illustrates that
${\cal P}_m(T)$ decays exponentially in the long-time limit.


\begin{figure}[h!]
\includegraphics[width=0.45\textwidth]{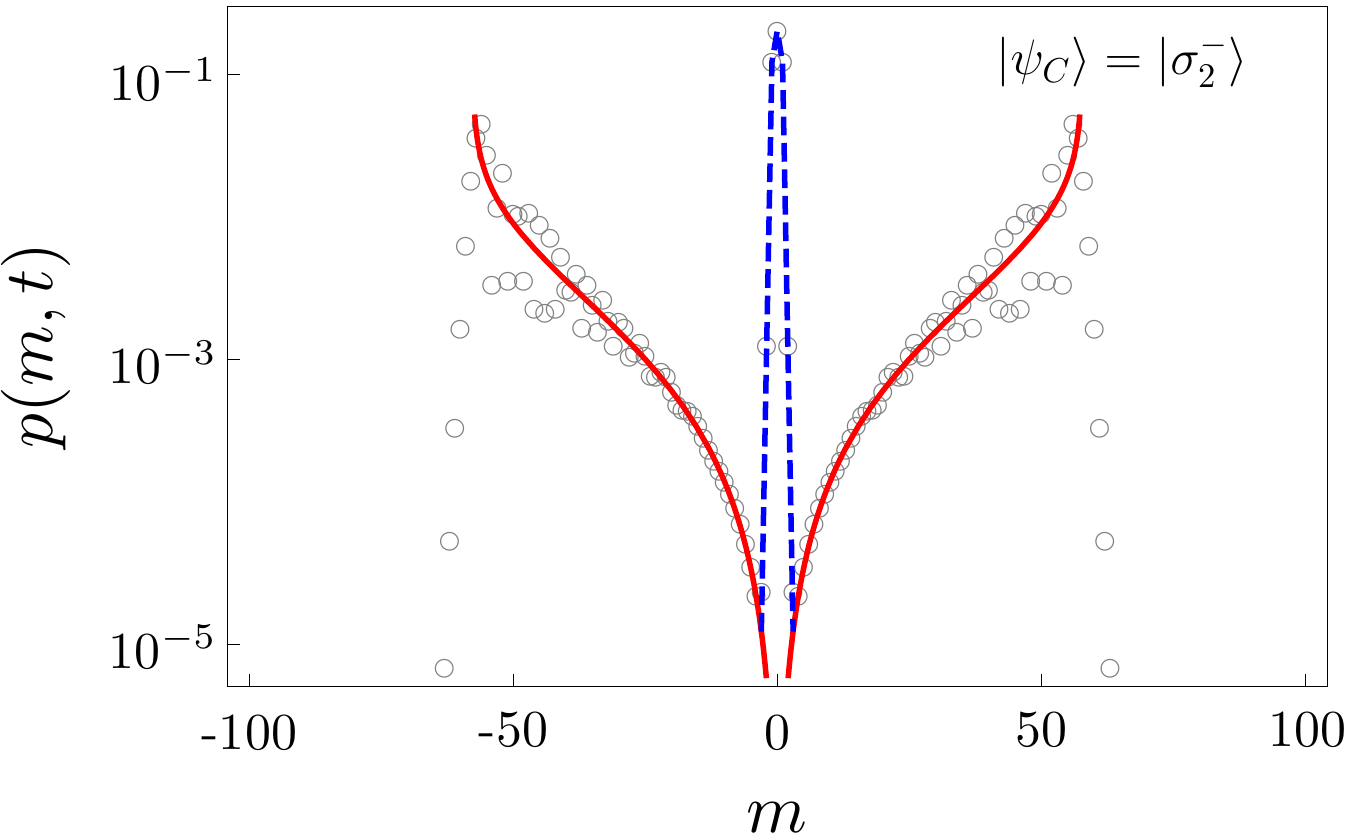}\vspace{12pt}
\includegraphics[width=0.45\textwidth]{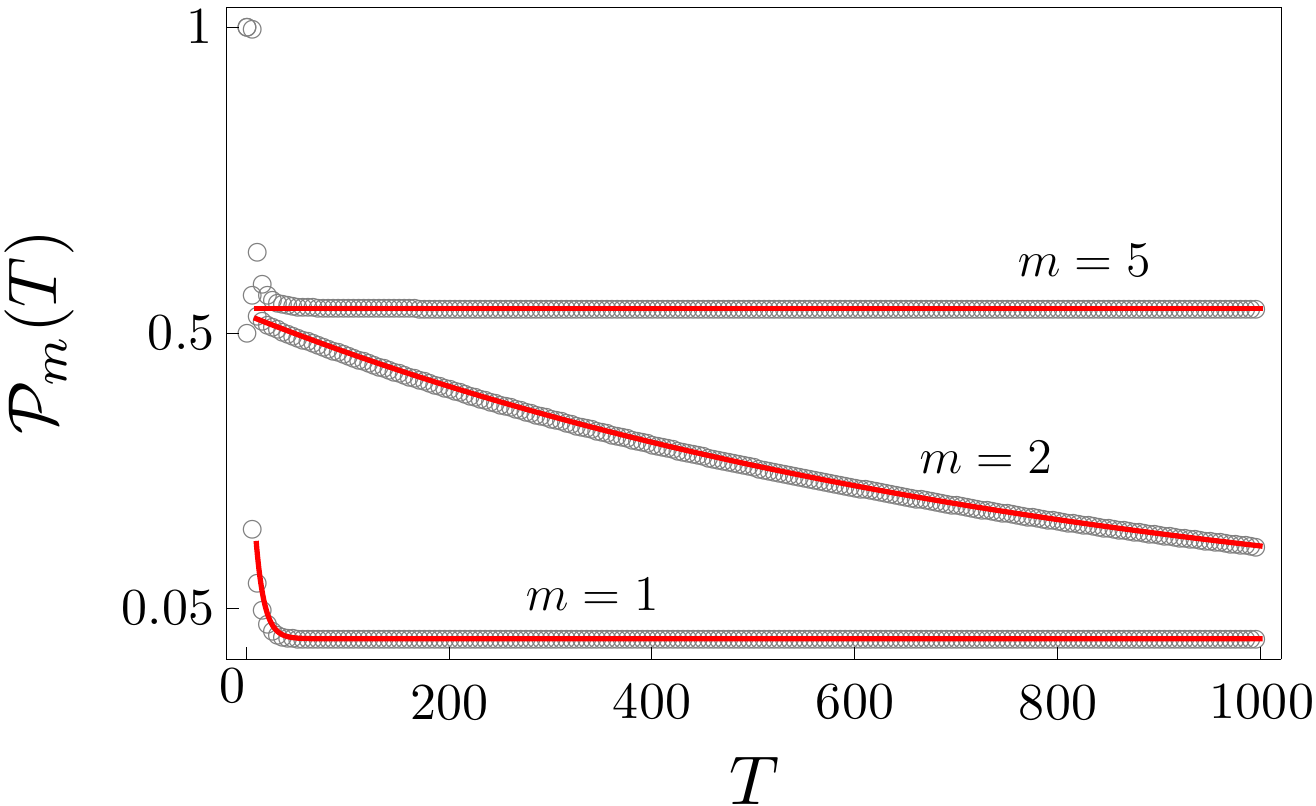}\vspace{12pt}
\includegraphics[width=0.45\textwidth]{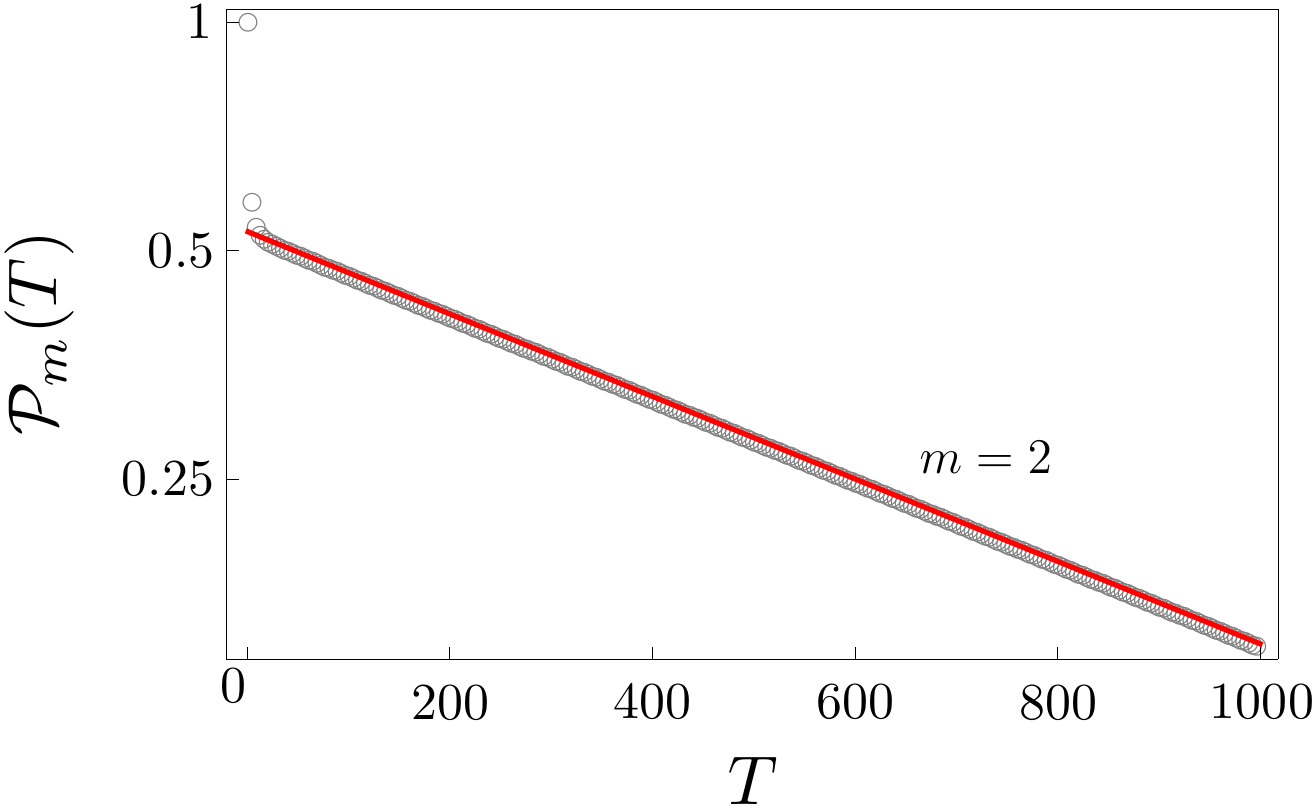}
\caption{Probability distribution and persistence for the three-state Grover walk starting with the coin state $|\psi_C\rangle = |\sigma_2^-\rangle$. In the upper plot we show the probability distribution after $t=100$ steps. The middle plot displays persistence (\ref{pers:exp}) of sites $m=1$, $m=2$ and $m=5$. The decay is exponential but the rate drops down very fast with the growing distance from the origin. The lower plot with the log-scale on the $y$-axis illustrates that the decay of persistence is indeed purely exponential (\ref{pers:exp}).}
\label{fig4}
\end{figure}

Let us now turn to the initial coin state $|\psi_C\rangle =
|\sigma_2^-\rangle$.  The general formula for persistence of site
$m$ (\ref{pers:3state}) for $g_2=1$ reduces into purely exponential
decay
\begin{equation}
\label{pers:exp}
{\cal P}_m^{(g_2)}(T) \sim e^{-\gamma(m) T},
\end{equation}
where the decay rate $\gamma(m)$ is given by (\ref{decay:const:gp}).
To illustrate this effect, we display in Figure~\ref{fig4} the
probability distribution and persistence for the Grover walk, i.e.
$\rho=1/\sqrt{3}$. The first plot shows the probability distribution
after $t=100$ steps. The second plot displays persistence of sites
$m=1$, $m=2$ and $m=5$. The decay rate (\ref{decay:const:gp})
decreases exponentially with the growing distance from the origin.
Hence, already for $m=5$ persistence essentially saturates on the considered time-scale. The last
plot shows persistence of site $m=2$ on a log-scale. The figure
illustrates that the decay of persistence is indeed purely
exponential.


\begin{figure}[h]
\includegraphics[width=0.45\textwidth]{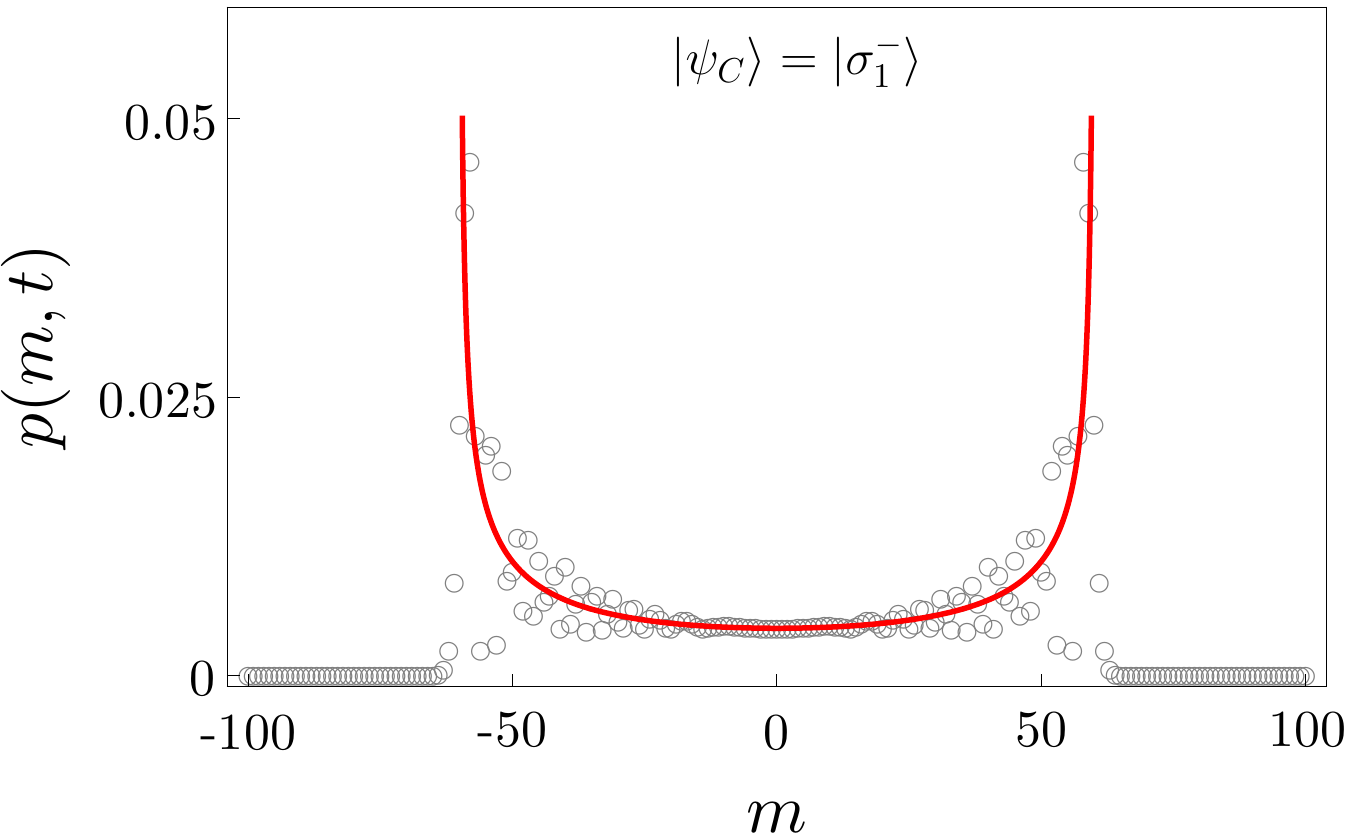}\vspace{12pt}
\includegraphics[width=0.45\textwidth]{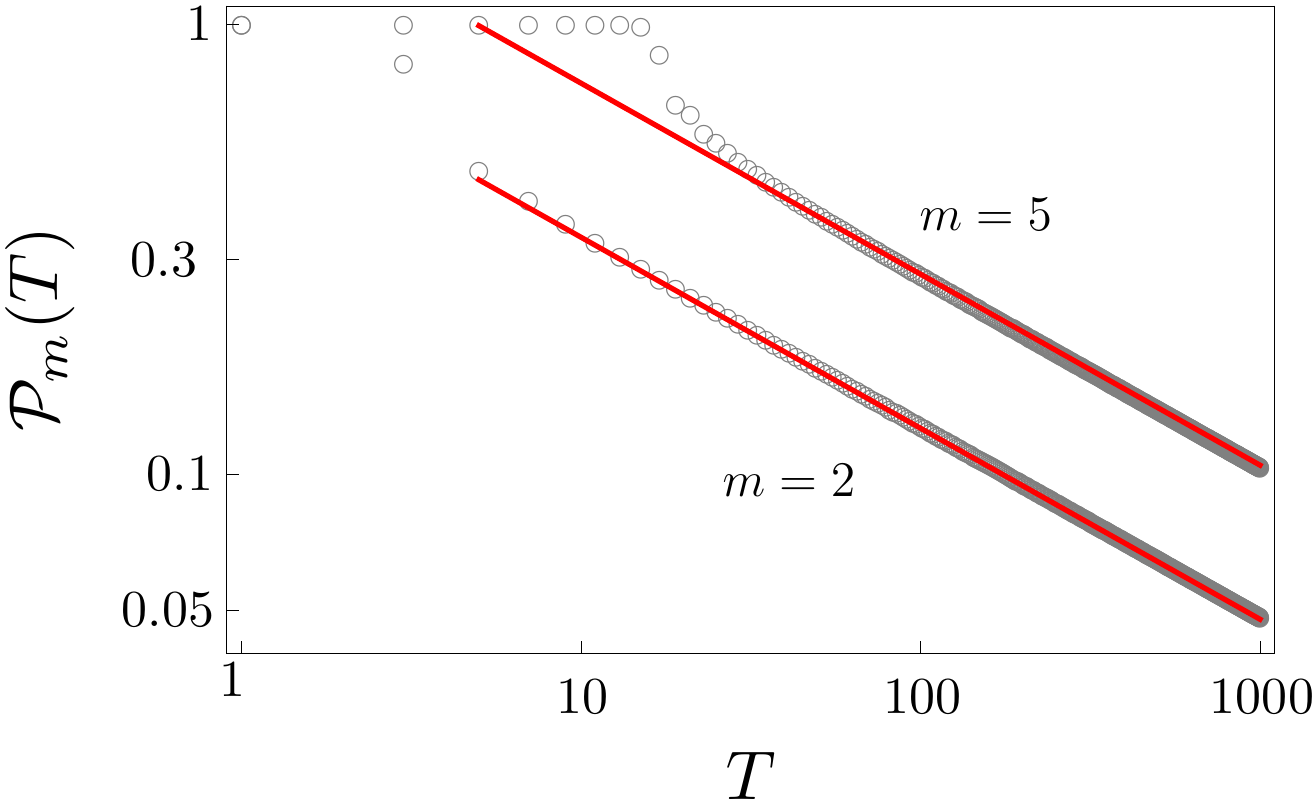}
\caption{Probability distribution and persistence for the three-state walk with $\rho=0.6$ starting with the coin state $|\psi_C\rangle = |\sigma_1^-\rangle$. The upper plot shows the probability distribution after $t=100$ steps. For this particular initial state the trapping effect disappears. The lower plot displays persistence of sites $m=2$ and $m=5$ on a log-log scale. We find that the scaling is given by the inverse power-law (\ref{pers:g1}).}
\label{fig5}
\end{figure}

Next, we consider the initial coin state $|\psi_C\rangle =
|\sigma_1^-\rangle$. In such a case the expression
(\ref{pers:3state}) reduces to a pure inverse power-law
\begin{equation}
\label{pers:g1}
{\cal P}_m^{(g_1)}(T) \sim \left(\frac{T}{|m|}\right)^{-\lambda},
\end{equation}
with the exponent $\lambda$ given by (\ref{power:decay}). To
illustrate this feature, we show in Figure~\ref{fig5} the
probability distribution and persistence for the three-state walk
with the coin parameter $\rho=0.6$. The upper plot displays the
probability distribution after 100 steps. We find that for the
particular initial state $|\psi_C\rangle = |\sigma_1^-\rangle$ the
trapping effect disappears. Indeed, according to (\ref{loc:rho}) we
find that $p_\infty(m)$ vanishes if $g_+ = g_2=0$. The lower plot
displays persistence of sites $m=2$ and $m=5$. The log-log scale
unravels that the scaling is given only by the inverse power-law
(\ref{pers:g1}).


Finally, let us point out that the dependence of the trapping
probability (\ref{loc:rho}) on the initial coin state can be
different for positive and negative $m$. This leads to different
behavior of persistence for sites on positive and negative
half-lines. As an example, consider the initial coin state
\begin{equation}
\label{psic:asym}
|\psi_C\rangle = \frac{1}{\sqrt{2}}\left(|\sigma^+\rangle
+|\sigma_2^-\rangle\right).
\end{equation}
We find that persistence of sites on positive half-line $(m>0)$ behaves like
\begin{equation}
\label{pers:pos:m}
{\cal P}_m^{+}(T) \sim \left(\frac{T}{m}\right)^{-\lambda} e^{-\gamma(m)T},
\end{equation}
where the exponent reads
\begin{equation}
\label{exp:pos:m}
\lambda = \frac{\sqrt{1-\rho^2}}{2\pi\rho},
\end{equation}
and the decay rate is given by
$$
\gamma(m) = \frac{4(1-\rho^2)}{\rho^4}Q^{2m}.
$$
Hence, for positive $m$ persistence decays exponentially in the
asymptotic regime. However, for sites on the negative half-line
$(m<0)$ persistence obeys only the inverse power-law
\begin{equation}
\label{pers:neg:m}
{\cal P}_m^{-}(T) \sim \left(\frac{T}{|m|}\right)^{-\lambda},
\end{equation}
with the exponent $\lambda$ given by (\ref{exp:pos:m}). We point out that coherence of the initial coin state is crucial for this effect. Indeed, consider the initial coin state given by an incoherent mixture of the basis states
$$
\rho_C = \frac{1}{2}|\sigma^+\rangle\langle\sigma^+| + \frac{1}{2}|\sigma_2^-\rangle\langle\sigma_2^-|.
$$
In such a case persistence is given by the sum of the expressions (\ref{pers:exp}) and (\ref{pers:g1}) with the corresponding exponent (\ref{power:decay}) and
decay rate (\ref{decay:const:gp}), independent of the sign of the position $m$. Hence, there is no asymmetry between negative and positive $m$ and persistence of all lattice sites decays exponentially in the asymptotic regime. Compared to the coherent superposition (\ref{psic:asym}) the exponent (\ref{power:decay}) is larger by a factor of two while the decay rate (\ref{decay:const:gp}) is smaller by a factor of two.

\begin{figure}[h!]
\includegraphics[width=0.45\textwidth]{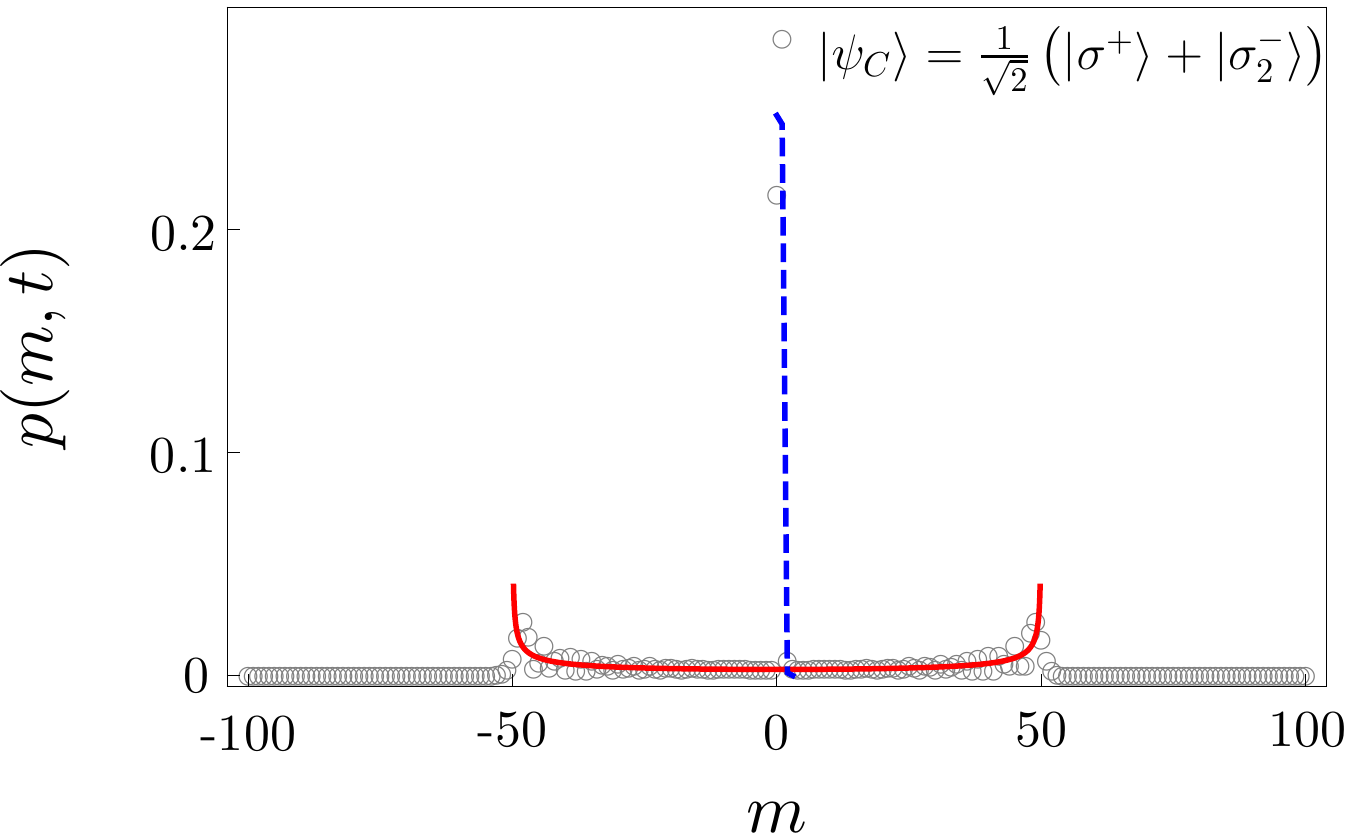}\vspace{12pt}
\includegraphics[width=0.45\textwidth]{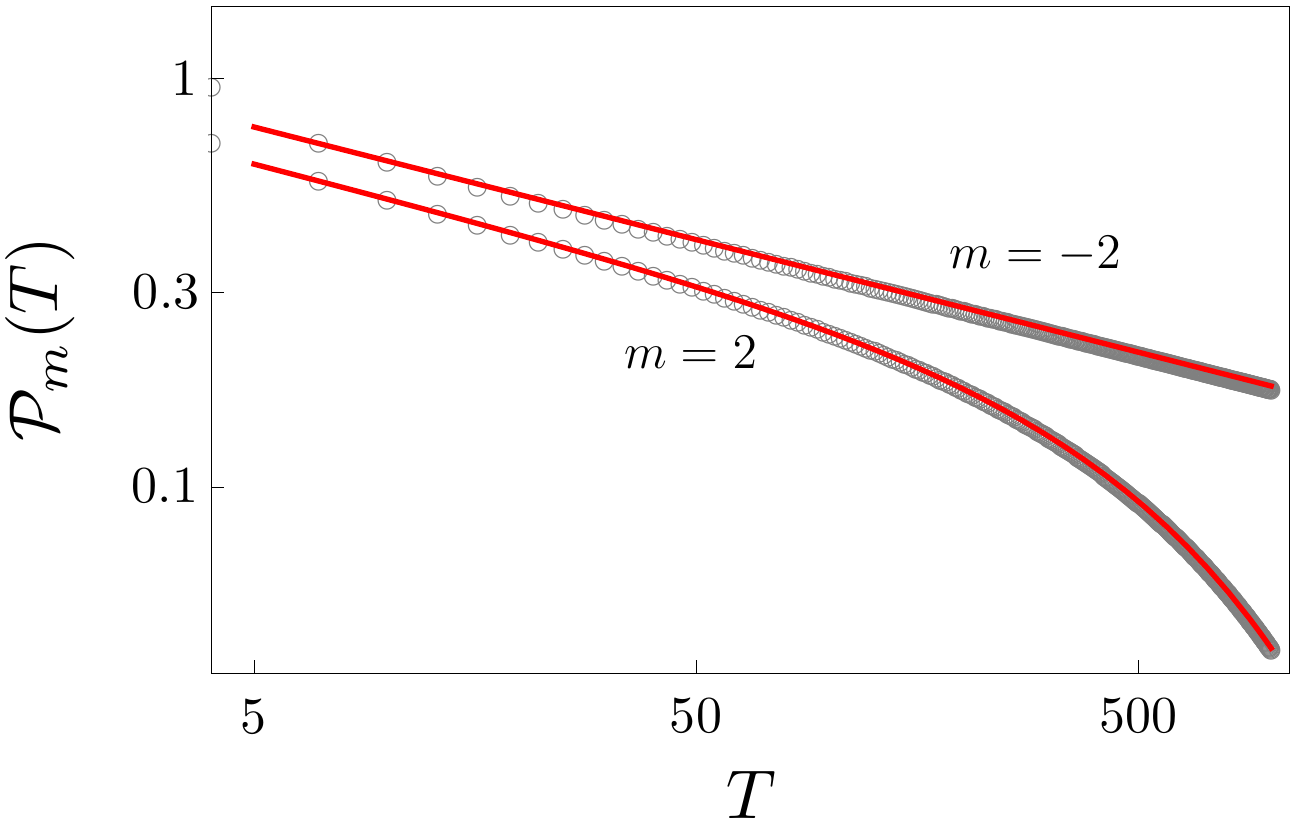}\vspace{12pt}
\includegraphics[width=0.45\textwidth]{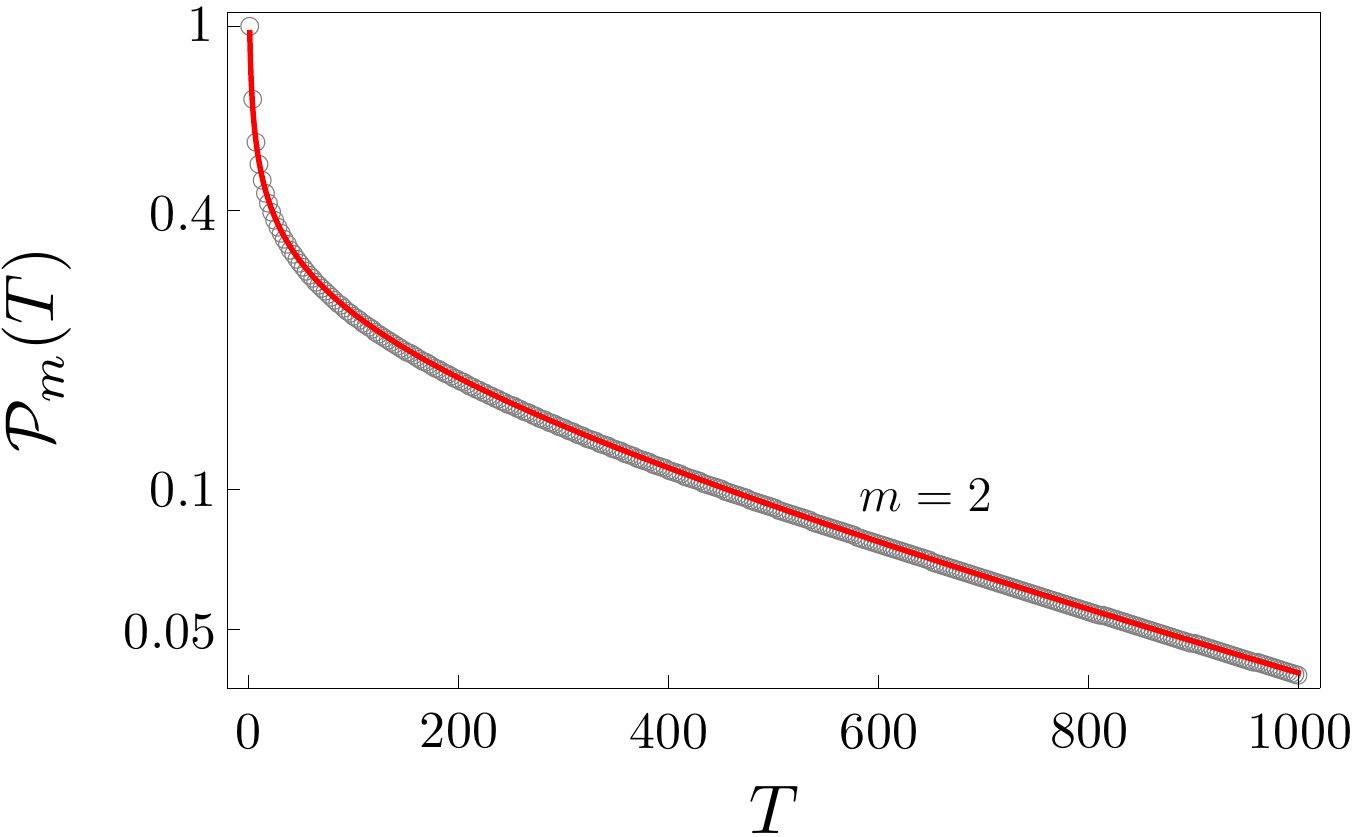}
\caption{Probability distribution and persistence for the three-state walk starting with the coin state (\ref{psic:asym}). The coin parameter was chosen as $\rho=0.5$. In the upper plot we display the probability distribution after $t=100$ steps. Note that the trapping probability is non-zero only on the positive half-line. The lower plot shows persistence of sites $m=2$ and $m=-2$ on the log-log scale. For $m=-2$ the behavior of persistence is determined only by the inverse power-law (\ref{pers:neg:m}). However, for $m=2$ the decrease of persistence is exponential (\ref{pers:pos:m}), as we illustrate in the last plot with the logarithmic scale.}
\label{fig6}
\end{figure}

We illustrate the results for the initial coin state (\ref{psic:asym}) in Figure~\ref{fig6} where we consider
the three-state quantum walk with the coin parameter $\rho=0.5$. In
the upper plot we display the probability distribution after 100
steps of the walk. Notice that the trapping probability, highlighted
by the dashed blue curve, is non-zero only on the positive
half-line. The lower plot illustrates the difference in the scaling
of persistence for sites on the positive or negative half-lines.
Here we show persistence of sites $m=2$ and $m=-2$ on the log-log
scale. We find that for $m=-2$ the behavior of persistence is
determined only by the inverse power-law (\ref{pers:neg:m}). On the
other hand, for $m=2$ the decrease of persistence is faster. Indeed,
for positive $m$ the behavior of persistence is dominated by the
exponential decay (\ref{pers:pos:m}) in the long-time limit. This is
illustrated in the last plot, where we show persistence of site
$m=2$ on the log-scale.


\section{Conclusions}
\label{sec:5}

In the present paper persistence of unvisited sites for two- and
three-state quantum walks on a line was analyzed. We have found that
in contrast to the classical random walk there is no connection
between the asymptotic behavior of persistence and scaling of the
variance with the number of steps. Concerning the two-state walk, we
have analytically confirmed the numerical result obtained in
\cite{pers:goswani} for the Hadamard walk. Moreover, we have
extended the analysis to a one-parameter set of two-state quantum
walks. In particular, we have shown that persistence of unvisited
sites obeys an inverse power-law independent of the initial
condition and the actual position of the site. The exponent of the
inverse power-law is determined by the parameter of the coin
operator.

The main result of the paper is the behaviour of persistence for
three-state quantum walks. We have focused on a one-parameter family
of walks which includes the familiar three-state Grover walk. Due to
the trapping effect displayed by the considered set of quantum
walks, the behaviour of persistence is more involved than for the
two-state quantum walks. In particular, we have shown that the
asymptotic scaling of persistence is in general determined by a
combination of an inverse power-law and an exponential decay.
However, both the exponent of the inverse power-law and the decay
rate of the exponential decline depend on the initial coin state.
Therefore, it is possible to obtain various asymptotic regimes of
persistence by choosing proper initial conditions. Moreover, one can
employ the asymmetry of the trapping effect to achieve different
asymptotic scaling of persistence for sites on the positive and
negative half-line. All obtained results have been facilitated by using a suitable basis formed by the eigenvectors of the coin operator. This allows to express persistence in closed and compact form and trace back the ways it is influenced
by the initial state and its coherence.

The present study is limited to the quantum walks on a line. A
natural extension is to consider persistence of unvisited sites in
quantum walks on higher-dimensional lattices. It would be
interesting if similar effects, such as the dependency of
persistence on the initial condition and various regimes of
persistence for different lattice sites, can be found on more
complicated lattices.


\begin{acknowledgments}

We appreciate the financial support from RVO~68407700. M\v S is
grateful for the financial support from GA\v CR under Grant No.
14-02901P. IJ is grateful for the financial support from GA\v CR
under Grant No. 13-33906S.

\end{acknowledgments}


\appendix


\section{Integral ${\cal I}_m(T)$ for a two-state walk}
\label{app:a}

We dedicate this appendix to evaluating integral ${\cal I}_m(T)$ defined in (\ref{pers:I}) for the two-state walk.  The limit density $w(v)$ is given by the formula (\ref{had:asymp:dist}). Since the limit density (\ref{had:asymp:dist}) is non-zero only for $|v|\leq \rho$, we replace the lower bound in the integral (\ref{pers:I}) with $|m|/\rho$. With the substitution $u = \frac{m}{\rho t}$ we rewrite ${\cal I}_m(T)$ into the form
\begin{widetext}
$$
{\cal I}_m(T) = \frac{\sqrt{1-\rho^2}}{\rho\pi}\int\limits_\frac{|m|}{\rho T}^1 \frac{1- {\rm sgn}(m) u(2|h_+|^2-1)}{u\left(1-\rho^2 u^2\right)\sqrt{1-u^2}} du.
$$
Evaluating the integral we obtain
\begin{eqnarray}
\nonumber {\cal I}_m(T) & = & \frac{\sqrt{1-\rho^2}}{\rho\pi} \ln\left(\frac{\rho T}{|m|} \left(1+\sqrt{1-\frac{m^2}{\rho^2 T^2}}\right)\right) + \frac{1}{\pi  }\arctan\left(\frac{\rho}{\sqrt{1-\rho^2}}\sqrt{1-\frac{m^2}{\rho^2 T^2}}\right) + \\
\nonumber & & + {\rm sgn}(m)\frac{2|h_+|^2-1}{\rho}\left(\frac{1}{\pi} \arctan\left(\frac{|m|}{\rho T} \sqrt{\frac{1-\rho^2}{1-\frac{m^2}{\rho^2 T^2}}}\right)-\frac{1}{2}\right).
\end{eqnarray}
Moreover, for large number of steps $T$ this function tends to
$$
{\cal I}_m(T) \approx \frac{\sqrt{1-\rho^2}}{\rho\pi}\ln\left(\frac{2\rho T}{|m|}\right) -\frac{\arcsin\rho}{\pi}+{\rm sgn}(m)\frac{2|h_+|^2-1}{2\rho}.
$$
Therefore, for large $T$ the function ${\cal I}_m(T)$ grows like a logarithm
$$
{\cal I}_m(T) \sim \lambda\ln\left(\frac{T}{|m|}\right),
$$
where the pre-factor reads
$$
\lambda = \frac{\sqrt{1-\rho^2}}{\rho\pi}.
$$


\section{Integral ${\cal I}_m(T)$ for a three-state walk}
\label{app:b}

In this appendix we evaluate the integral (\ref{pers:I}) for a three-state quantum walk, i.e. the limit density is given by (\ref{dist:rho}). Using the substitution $u = \frac{m}{\rho t}$ we rewrite ${\cal I}_m(T)$ into the form
$$
{\cal I}_m(T) = \frac{\sqrt{1-\rho^2}}{\rho \pi}\int\limits_\frac{|m|}{\rho T}^1 \frac{1-|g_2|^2-(g_1\overline{g_2} + \overline{g_1}g_2)u + (|g_2|^2 - |g_+|^2)u^2}{u(1-u^2)\sqrt{1-u^2}} du.
$$
Evaluating the integral we obtain the following result
\begin{eqnarray}
\nonumber {\cal I}_m(T) & = & \frac{\sqrt{1-\rho^2}}{\pi  \rho} \left(1-\left| g_2\right| ^2\right) \left(\ln \left(\frac{\rho T}{|m|}\right) + \ln \left(1+ \sqrt{1-\frac{m^2}{\rho^2 T^2}}\right)\right) - \\
\nonumber & & -\frac{1}{2\pi}\left(1-\left| g_2\right| ^2\right) \arctan\left(\frac{2 \rho \sqrt{\left(1-\frac{m^2}{\rho^2 T^2}\right) \left(1-\rho^2\right)}}{\left(2-\frac{m^2}{\rho^2 T^2}\right) \rho^2-1}\right)  +\frac{1}{\pi  \rho^2}\left(\left| g_2\right| ^2-\left| g_+\right| ^2\right) \arctan\left(\rho \sqrt{\frac{1-\frac{m^2}{\rho^2 T^2}}{1-\rho^2}}\right) - \\
\nonumber & & -\frac{1}{2\pi\rho}\left(\overline{g_1}g_2 + g_1 \overline{g_2}\right) \left(\pi-2\arctan\left(\frac{\frac{|m|}{\rho T} \sqrt{1-\rho^2}}{\sqrt{1-\frac{m^2}{\rho^2 T^2}}}\right)\right).
\end{eqnarray}
For large number of steps $T$ this function approaches
\begin{eqnarray}
\nonumber {\cal I}_m(T) & \approx & \frac{\sqrt{1-\rho^2}}{\pi  \rho} \left(1-\left| g_2\right| ^2\right) \ln \left(\frac{2\rho T}{|m|}\right) + \frac{1}{2\pi}\left(1-\left| g_2\right| ^2\right) \arctan\left(\frac{2 \rho \sqrt{1-\rho^2}}{1-2\rho^2}\right) + \\
\nonumber & & + \frac{1}{\pi  \rho^2}\left(\left| g_2\right| ^2-\left| g_+\right| ^2\right) \arctan\left(\rho \sqrt{\frac{1}{1-\rho^2}}\right) - \frac{1}{2\rho}\left(\overline{g_1}g_2 + g_1 \overline{g_2}\right)
\end{eqnarray}
Wee see that ${\cal I}_m(T)$ asymptotically grows like a logarithm
$$
{\cal I}_m(T) \sim \lambda\ln\left(\frac{T}{|m|}\right),
$$
where the pre-factor reads
$$
\lambda = \frac{\sqrt{1-\rho^2}}{\pi  \rho} \left(1-\left| g_2\right| ^2\right).
$$
\end{widetext}


\end{document}